\documentclass[preprint,showpacs,preprintnumbers,amsmath,superscriptaddress,amssymb,10pt]{revtex4}
\usepackage{graphicx}% Include figure files
\usepackage{dcolumn}% Align table columns on decimal point
\usepackage{bm}% bold math
\usepackage{amssymb}
\usepackage{color}

\def\bea{\begin{eqnarray}}
\def\ena{\end{eqnarray}}

\begin{document}

%%%%%%%%%%%%%%%%%%%%%%%%%%%%%%%%%%%%%%%%%%%%%%%%%%%%%%%%%%%%%%%%%
\title{Response of a Spaceborn Gravitational Wave Antenna to Solar Oscillations}
%%%%%%%%%%%%%%%%%%%%%%%%%%%%%%%%%%%%%%%%%%%%%%%%%%%%%%%%%%%%%%%%%

%%%%%%%%%%%%%%%%%%%%%%%%%%%%%%%%%%%%%%%%%%%%%%%%%%%%%%%%%%%%%%%%%

\author{A.~G.~Polnarev}
\email{A.G.Polnarev@qmul.ac.uk}
\affiliation{ Astronomy Unit, School of Mathematical Sciences
Queen Mary, University of London, London E1 4NS, UK}
\author{I.~W.~Roxburgh}
\affiliation{ Astronomy Unit, School of Mathematical Sciences
Queen Mary, University of London, London E1 4NS, UK}
\affiliation{LESIA, Observatoire de Paris, 92155 Meudon, France}
\author{D.~Baskaran}
\email{Deepak.Baskaran@astro.cf.ac.uk}
\affiliation{School of Physics and Astronomy, Cardiff University,
Cardiff CF24 3AA, UK}

%%%%%%%%%%%%%%%%%%%%%%%%%%%%%%%%%%%%%%%%%%%%%%%%%%%%%%%%%%%%%%%%%

\small

%%%%%%%%%%%%%%%%%%%%%%%%%%%%%%%%%%%%%%%%%%%%%%%%%%%%%%%%%%%%%%%%%

\begin{abstract}
{We investigate the possibility of observing very small amplitude low frequency solar oscillations with the proposed laser interferometer space antenna (LISA).  For frequencies $\nu$ below $3\times 10^{-4}~{\rm Hz}$ the dominant contribution is from the near zone time dependent gravitational quadrupole moments associated with the normal modes of oscillation. For frequencies $\nu$ above $ 3\times 10^{-4}~{\rm Hz}$ the dominant contribution is from gravitational radiation generated by the quadrupole oscillations which is larger than the Newtonian signal by a factor of the order $(2 \pi r \nu/ c)^4$, where $r$ is the distance to the Sun, and $c$ is the velocity of light.

The low order solar quadrupole pressure and gravity oscillation modes have not yet been detected above the solar background by helioseismic velocity and intensity measurements. We show that for frequencies $\nu \lesssim 2\times 10^{-4}~{\rm Hz}$, the signal due to solar oscillations will have a higher signal to noise ratio in a LISA type space interferometer than in helioseismology measurements. Our estimates of the amplitudes needed to give a detectable signal on a LISA type space laser interferometer imply surface velocity amplitudes on the sun of the order of $1-10$ mm/sec in the frequency range $1\times 10^{-4} -5\times 10^{-4}~{\rm Hz}$. If such modes exist with frequencies and amplitudes in this range they could be detected with a LISA type laser interferometer.}

\end{abstract}

%%%%%%%%%%%%%%%%%%%%%%%%%%%%%%%%%%%%%%%%%%%%%%%%%%%%%%%%%%%%%%%%%


%\today

\maketitle

%%%%%%%%%%%%%%%%%%%%%%%%%%%%%%%%%%%%%%%%%%%%%%%%%%%%%%%%%%%%%%%%%%%%%%%%%%%%%%%%%%%%%%%%
%%%%%%%%%%%%%%%%%%%%%%%%%%%%%%%%%%%%%%%%%%%%%%%%%%%%%%%%%%%%%%%%%%%%%%%%%%%%%%%%%%%%%%%%

\section{Introduction \label{SectionI}}

The proposed ESA/NASA gravitational wave laser interferometric space antenna (LISA) \cite{Bender1999,LISAWebsite} will give a unique window into gravitational wave physics. LISA will be sensitive to gravitational waves in the frequency range between $10^{-4}$ and 1 Hz, a range currently inaccessible on the ground due to seismic noise.  LISA will consist of three spacecraft at the vertices of an equilateral triangle of sides $5\times10^9$ metres. The system would be maintained in this configuration by arranging that the plane of the detectors has an inclination of $60^o$ to the ecliptic and counter rotates with the same period as it orbits the Sun. Any two arms will constitute a Michelson type interferometer, a mother spacecraft will send a laser beam to the other two satellites where the signal would be coherently transponded back. The interferometer readout would be obtained by interfering the incoming signal with the outgoing one and comparing the fractional change in phase shift between the two arms.

Current studies of the sensitivity of the LISA experiment \cite{Bender1999} indicate that the instrumental noise, due dominantly to the residual uncompensated accelerations, would be of the order $3\times 10^{-18} /\sqrt{\rm Hz}$ at $10^{-4}~{\rm Hz}$ (in units of dimensionless strain per root Hertz). In addition there is likely to be background ``confusion noise" from binary systems which has a comparable magnitude at $10^{-4}~{\rm Hz}$ \cite{Bender1997,Abbott2004,Barack2004}. With one year's observation the dimensionless strain that could be detected at a signal to noise ratio of 5 is estimated to be around $10^{-21} - 10^{-20}$.

The primary goal of LISA is to detect gravitational waves from individual sources (close binary systems, neutron star or black hole coalescence) and any stochastic background due to the superposition of waves emitted by binary systems, and possibly from the early Universe \cite{Grishchuk2001,Cutler2002,Sathyaprakash2009}. Apart from the primary goal, in general, a LISA type interferometer is sensitive to any variations in the gravitational field in the frequency range $10^{-4} - 1~{\rm Hz}$.  Since the Sun is known to be oscillating in normal modes of small amplitude with frequencies in this range, if the amplitudes are large enough the oscillating external gravitational field could contribute to the signal detected by LISA \cite{Cutler1996,Polnarev1996,Roxburgh1999,Roxburgh2000} (previously, independently suggested in \cite{Schutz1995,Gough1995}).

At the present time solar oscillations can only be detected through surface variations in velocity and luminosity. The first evidence of surface layer solar oscillations dates back to the work \cite{Leighton1962}. The low order gobal oscillations, which are the oscillations of interest here, were detected by \cite{Claverie1979,Grec1980} as resolved peaks in the power spectrum of a time series of measurements of the Doppler shift of a K and Na line using the integrate light from the Sun. As a result of ground based and space based observational programmes upwards of $10^7$ oscillation modes have been identified in the frequency range $10^{-3} - 10^{-2}~{\rm Hz}$.; knowledge of these frequencies has been used to infer the acoustic and dynamical structure of the Sun (pressure, density and rotation as a function of radius), placing constraints on  the physics of the solar interior and on models of solar (and thereby stellar) evolution.

The relation between the surface amplitude in velocity and the magnitude of the oscillating gravitational field depends on a detailed understanding of the behaviour of the oscillations in the outer layers of the Sun which is not well understood. Such modes (g-modes and low order p-modes) are difficult to detect above the solar background noise, which increases at low frequencies. Great effort is currently being expended in the search for such modes. In the paper \cite{Garcia2007} the authors claim to have  detected a g-mode at a frequency of $1.5\times 10^{-4}~{\rm Hz}$ with an amplitude of the order $1~{\rm cm}$, but others have not detected any such modes at this amplitude \cite{Appourchaux2000}. Since only quadrupole modes could have amplitudes large enough to be detected by a gravitational wave detector at 1 a.u., we confine our study to such modes.

At a distance of $r=1~{\rm a.u.}$ these oscillations are in the near zone (Newtonian regime) for frequencies $\nu$ such that $c/2\pi\nu > r$ (i.e.~$\nu < \nu_r \approx 3\times10^{-4}~{\rm Hz}$). The observational implication of the near zone oscillations for LISA was studied in \cite{Cutler1996}. For larger frequencies, $1~{\rm a.u.}$ is already in the wave zone which means that in addition to the time dependent external Newtonian gravitational field the time varying gravitational quadrupole moments also generate gravitational waves, which could give a detectable signal on a LISA type interferometer for frequencies $\nu > \nu_r \sim 3\times10^{-4}~{\rm Hz}$ \cite{Polnarev1996,Roxburgh1999, Roxburgh2000}.

In the present work we investigate the possibility of detecting such low frequency quadrupole oscillations with a LISA type laser interferometer, including both the Newtonian near zone perturbations and the associated gravitational wave emission in the frequency range $3\times 10^{-5} - 10^{-3}~{\rm Hz}$. We compare the gravitational signals detectable by laser interferometry with the velocity signals detectable by whole disc helioseismolgy, and demonstrate that low frequency quadrupole oscillations with surface velocity amplitudes below current helioseimic limits could nevertheless be large enough to contribute to the signal detected by LISA.  If such modes are first detected by helioseismic techniques then the measured frequencies (and predicted power) would provide a valuable calibration tool for LISA. On the other hand, if they are not detected by helioseismic means then one can look upon LISA as a potential telescope for studying the deep solar interior.

The plan of this paper is as follows. In Section \ref{SectionII} we begin with a discussion of the properties of the solar oscillations and quantify the relationship between the surface radial amplitude of an oscillation, its quadrupole moment and its horizontal amplitude.  We express the external gravitational field in terms of the quadrupole moment tensor expressed as a sum over a set of basis tensors (corresponding to surface harmonics) which enables us to relate the generation of gravitational waves to the surface amplitude of the oscillation. In Section \ref{SectionIII} we derive the expected phase shift  in an interferometer arm due to both the time dependent Newtonian field and to the associated gravitational waves for a given oscillating quadrupole moment. These results are then used in Section \ref{SectionIV} to determine the response of LISA. Our analysis shows that above frequencies $\nu\sim 3\times 10^{-4}~{\rm Hz}$ the signal is dominated by gravitational waves. Next, we determine the magnitude of the velocity signal from each mode in terms of surface amplitude. In Section \ref{SectionV} in order to analyze the prospects of detectability, we consider the background noise for both velocity and gravitational detectors and construct the signal to noise for both gravity and velocity experiments for a given assumed frequency resolution. The ratio of these S/N is then independent of the assumed surface amplitude of the oscillation and if this ratio is greater than 1 the modes are easier to detect by a gravitational laser interferometer than by helioseismic experiments.  The outcome of these calculations is summarised in Figure \ref{figure4}. It follows that, for frequencies $\nu\lesssim 2\times 10^{-4}~{\rm Hz}$ the most of the quadrupole modes are more readily detected in by a LISA type interferometer than by helioseismic experiments. Finally, we present our conclusions in Section \ref{SectionVII}.

%%%%%%%%%%%%%%%%%%%%%%%%%%%%%%%%%%%%%%%%%%%%%%%%%%%%%%%%%%%%%%%%%%%%%%%%%%%%%%%%%%%%%%%%
%%%%%%%%%%%%%%%%%%%%%%%%%%%%%%%%%%%%%%%%%%%%%%%%%%%%%%%%%%%%%%%%%%%%%%%%%%%%%%%%%%%%%%%%

\section{Solar Oscillations and metric perturbation field around the Sun \label{SectionII}}

The solar oscillations are normally expressed in terms of a surface harmonic and Fourier time decomposition with any variable, for example the radial displacement ${\delta} {r}$, expressed in the form
\bea
{\delta {r}({\bf r},t)\over R_\odot} = \sum_{\ell=0}^\infty~\sum_{m=-\ell}^\ell~
\sum_{n=-\infty}^\infty {\bf\zeta}_{n\ell m}(r)\,S_{\ell m}
(\theta,\phi)\,e^{i\omega t},
\label{1}
\ena
where $\omega = \omega_{n\ell m}$ are the cyclic eigenfrequencies of modes corresponding to a particular surface harmonic $S_{\ell m}(\theta,\phi)$, $n$ labels the order of the mode ($|n|$ is essentially the overtone number, the number of nodes in the radial direction), $\zeta_{n\ell m}(r)$ the corresponding dimensionless eigenfunctions and $(r,\theta,\phi)$ spherical polar coordinates with origin at the centre of the Sun. The modes are classified as p (pressure) modes with frequencies increasing with increasing $n$, and g (gravity) modes with frequencies decreasing with increasing $n$, (for clarity we use negative $n$ for the g-modes in (\ref{1})). If the basic unperturbed state is spherically symmetric the frequencies are independent of azimuthal order $m$. Rotation lifts this degeneracy giving frequencies $\omega_{n\ell m} \approx \omega_{n\ell 0} + m\bar{\Omega}$ where $\bar{\Omega}\sim\Omega_\odot\sim 3\times10^{-6}$ rad/sec is a weighted mean of the solar angular velocity. With a frequency resolution of $\Delta\nu \sim 3\times 10^{-8}~{\rm Hz}$, as envisaged for LISA (with 1 year's observation time), these individual m-value modes should be resolved provided there is sufficient power and the line widths are sufficiently narrow. For frequencies $\nu = \omega/2\pi \sim 10^{-4}~{\rm Hz}$, which is the region of interest in the present analysis,  $\Omega_\odot/\omega \sim 10^{-3}$ and the eigenfunctions of the modes may be taken to be independent of azimuthal order $m$. The solar rotation axis is inclined at an angle of about $7^o$ to the ecliptic plane, which introduces an additional (small) modulation which is  neglected in the present analysis.

There is currently some debate over whether or not any oscillations with frequencies $\nu\sim 10^{-4}~{\rm Hz}$ have yet been detected ({\it cf.}~\cite{Garcia2007}).  The measured line widths at higher frequencies decrease with decreasing frequency and crude extrapolation from the measured range suggests $\Delta\nu < 3\times 10^{-8}~{\rm Hz}$ at frequencies $\nu\sim 10^{-4}~{\rm Hz}$ and consequently mode lifetimes more than $1$ year.  We shall assume here that the lines are narrower than the frequency resolution and the modes may effectively be considered as monochromatic. Since the external gravitational potential of a multipole  of order $\ell$ decreases like $1/r^{\ell+1}$ the modes with $\ell=2$ will dominate at $1~{\rm a.u.}$. For this reason we shall only consider the quadrupole modes $\ell = 2$ in the present work. Furthermore, since the oscillation velocities are very small compared with the velocity of light only quadrupole gravitational radiation will be significant.

The external Newtonian gravitational potential of the oscillating Sun can then be expressed in the equivalent forms
\bea
U({\bf r},t) =  -G\,\int_\odot {\rho dV\over r}
 = U_0({\bf r})
-{G\over 6}\,{\cal D}^{\alpha\beta}\,
\nabla_{\alpha\beta}\left(1\over r\right)
= U_0({\bf r}) - G M_\odot R^2_\odot  \sum_{m=-2}^2~\sum_{n=-\infty}^\infty
{J_{nm}\over r^3} S_{2 m}(\theta, \phi),
\label{2}
\ena
where $U_0({\bf r})$ is the time independent potential, $(r,\theta,\phi$) are spherical polar coordinates, $S_{2 m}(\theta,\phi)$ surface harmonics of degree $\ell = 2$ (normalised to unity over a sphere), $J_{nm}\propto e^{i\omega t}$ the dimensionless time dependent quadrupole moments corresponding to eigenmodes with cyclical frequencies $\omega = \omega_{nm}$. In the above expression ${\cal D}^{\alpha\beta}$ is the quadrupole moment given by
\bea
{\cal D}^{\alpha\beta} = \int_\odot \left(3 x^\alpha x^\beta
- \delta^{\alpha\beta} x^\mu x_\mu\right)\rho(t)\,dV
 = M_\odot R^2_\odot\sum_{n=-\infty}^\infty\,
\sum_{m=-2}^2 C_m\,J_{nm}{\cal I}_{m}^{\alpha\beta},
\label{3}
\ena
where, $x^\alpha$ correspond to a Cartesian coordinate system (with $x^3$ along the $\theta=0$ and $x^1$ along $\phi=0$, $\theta=\pi/2$ axes), $\delta^{\alpha\beta}$ the Kronecker delta, $\nabla_{\alpha\beta} = \partial^2/\partial x^\alpha \partial x^\beta$, and ${\cal I}_m^{\alpha\beta}$ are the set of trace-free basis tensors corresponding to surface harmonics and $C_{0} = -\sqrt{5/4\pi},~C_{m} = \sqrt{15/4\pi}~({\rm for}~m\ne 0)$. We refer the reader to Appendix \ref{AppendixI} for details.

\begin{figure}
\begin{center}
\includegraphics[width=12cm]{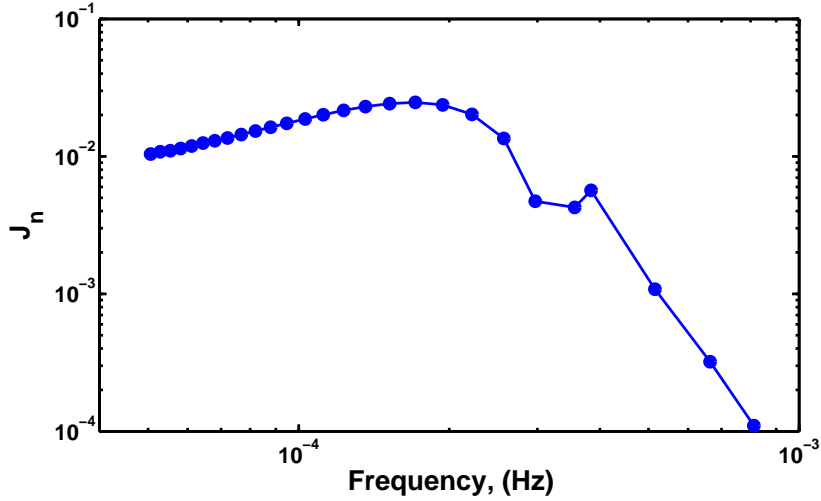}
\end{center}
\caption{Normalised quadrupole moments $J_{n}$ (in units of $G M R^2_\odot$) for quadrupolar solar oscillations with surface displacement $\zeta_n(R)=1$.}\label{figure1}
\end{figure}

The properties of the eigenmodes were computed using the standard Aarhus Solar Model S1 \cite{Christensen-Dalsgaard1996}, with surface radial velocity amplitudes normalised to $1~{\rm cm/sec}$  at the solar surface $r=R_\odot$. The amplitude of oscillating quadrupole moments $J_n$ are given in Figure \ref{figure1} (we have suppressed the subscript $\ell = 2$ and $m$, since the eigensolutions are independent of $m$). Table \ref{table1}  gives a summary of the results obtained for the amplitudes of relevant quantities derived in this and subsequent sections. Column 1 gives the radial order $n$ of the mode, column 2 the frequency $\nu$, column 3 the horizontal displacement eigenfunction at the solar surface $\zeta_h$, column 4 the quadrupole moment $J$ and column 5 gives the total kinetic energy of the mode $E$.

\begin{table}
\begin{center}
\begin{tabular}{|c|c|c|c|c|c|c|c|c|c|c|c|c|}
\hline
{\tiny 1} & {\tiny 2}  & {\tiny 3}  & {\tiny 4}  & {\tiny 5}  & {\tiny 6}  & {\tiny 7}  & {\tiny 8}  & {\tiny 9}  & {\tiny 10}  & {\tiny 11}  & {\tiny 12} & {\tiny 13}    \\
\hline
& & & & & & & & & & &
\\
$~~n~~$ & $\nu$ & $\zeta_h$ & $J$ & $E$  & $S_0$ & $S_1$ & $S_2$ & $V_0$ & $V_2$ & $B_I$ & $B_b$ & $B_v$ \\
& & & & & & & & & & & 
\\
\hline\hline
-22 & 5.06E-05 & 3.86E+00 & 1.04E-02 & 1.31E+50 & 1.28E-14 & 1.62E-13 & 2.12E-13 & 2.32E+05 & 2.85E+05 & 1.19E-17 & 1.08E-17 & 2.89E+01  \\
\hline
-21 & 5.29E-05 & 3.54E+00 & 1.08E-02 & 1.13E+50 & 1.21E-14 & 1.53E-13 & 2.01E-13 & 2.24E+05 & 2.75E+05 & 1.09E-17 & 1.02E-17 & 2.85E+01  \\
\hline
-20 & 5.54E-05 & 3.23E+00 & 1.10E-02 & 9.35E+49 & 1.13E-14 & 1.43E-13 & 1.87E-13 & 2.16E+05 & 2.65E+05 & 9.91E-18 & 9.53E-18 & 2.77E+01  \\
\hline
-19 & 5.81E-05 & 2.93E+00 & 1.14E-02 & 7.93E+49 & 1.07E-14 & 1.35E-13 & 1.77E-13 & 2.08E+05 & 2.55E+05 & 9.01E-18 & 8.88E-18 & 2.69E+01  \\
\hline
-18 & 6.11E-05 & 2.65E+00 & 1.19E-02 & 6.67E+49 & 1.01E-14 & 1.27E-13 & 1.67E-13 & 2.00E+05 & 2.45E+05 & 8.15E-18 & 8.19E-18 & 2.65E+01  \\
\hline
-17 & 6.44E-05 & 2.38E+00 & 1.25E-02 & 5.53E+49 & 9.49E-15 & 1.20E-13 & 1.57E-13 & 1.93E+05 & 2.36E+05 & 7.34E-18 & 7.52E-18 & 2.59E+01  \\
\hline
-16 & 6.80E-05 & 2.13E+00 & 1.30E-02 & 4.50E+49 & 8.87E-15 & 1.12E-13 & 1.46E-13 & 1.85E+05 & 2.27E+05 & 6.57E-18 & 6.94E-18 & 2.40E+01  \\
\hline
-15 & 7.21E-05 & 1.89E+00 & 1.36E-02 & 3.67E+49 & 8.30E-15 & 1.04E-13 & 1.37E-13 & 1.78E+05 & 2.18E+05 & 5.84E-18 & 6.29E-18 & 2.45E+01  \\
\hline
-14 & 7.68E-05 & 1.67E+00 & 1.44E-02 & 2.95E+49 & 7.76E-15 & 9.74E-14 & 1.28E-13 & 1.71E+05 & 2.09E+05 & 5.16E-18 & 5.53E-18 & 2.35E+01  \\
\hline
-13 & 8.20E-05 & 1.46E+00 & 1.53E-02 & 2.34E+49 & 7.23E-15 & 9.05E-14 & 1.19E-13 & 1.64E+05 & 2.01E+05 & 4.52E-18 & 4.77E-18 & 2.23E+01  \\
\hline
-12 & 8.79E-05 & 1.27E+00 & 1.63E-02 & 1.82E+49 & 6.72E-15 & 8.39E-14 & 1.10E-13 & 1.58E+05 & 1.93E+05 & 3.93E-18 & 4.01E-18 & 2.21E+01  \\
\hline
-11 & 9.47E-05 & 1.09E+00 & 1.74E-02 & 1.39E+49 & 6.21E-15 & 7.72E-14 & 1.01E-13 & 1.52E+05 & 1.86E+05 & 3.39E-18 & 3.57E-18 & 2.18E+01  \\
\hline
-10 & 1.03E-04 & 9.30E-01 & 1.87E-02 & 1.05E+49 & 5.72E-15 & 7.06E-14 & 9.25E-14 & 1.46E+05 & 1.79E+05 & 2.89E-18 & 3.25E-18 & 2.13E+01  \\
\hline
  -9 & 1.12E-04 & 7.81E-01 & 2.01E-02 & 7.77E+48 & 5.23E-15 & 6.40E-14 & 8.39E-14 & 1.41E+05 & 1.73E+05 & 2.43E-18 & 3.06E-18 & 2.05E+01  \\
\hline
  -8 & 1.23E-04 & 6.47E-01 & 2.16E-02 & 5.52E+48 & 4.73E-15 & 5.70E-14 & 7.47E-14 & 1.37E+05 & 1.68E+05 & 2.02E-18 & 2.83E-18 & 2.06E+01  \\
\hline
  -7 & 1.36E-04 & 5.29E-01 & 2.30E-02 & 3.86E+48 & 4.23E-15 & 4.98E-14 & 6.52E-14 & 1.35E+05 & 1.65E+05 & 1.65E-18 & 2.65E-18 & 1.89E+01  \\
\hline
  -6 & 1.52E-04 & 4.24E-01 & 2.42E-02 & 2.59E+48 & 3.71E-15 & 4.21E-14 & 5.51E-14 & 1.33E+05 & 1.63E+05 & 1.33E-18 & 2.48E-18 & 1.87E+01  \\
\hline
  -5 & 1.71E-04 & 3.33E-01 & 2.47E-02 & 1.62E+48 & 3.18E-15 & 3.39E-14 & 4.43E-14 & 1.33E+05 & 1.63E+05 & 1.04E-18 & 2.34E-18 & 1.76E+01  \\
\hline
  -4 & 1.94E-04 & 2.57E-01 & 2.37E-02 & 9.12E+47 & 2.61E-15 & 2.52E-14 & 3.28E-14 & 1.36E+05 & 1.66E+05 & 8.05E-19 & 2.22E-18 & 1.68E+01  \\
\hline
  -3 & 2.22E-04 & 1.97E-01 & 2.02E-02 & 4.87E+47 & 1.99E-15 & 1.65E-14 & 2.15E-14 & 1.41E+05 & 1.73E+05 & 6.15E-19 & 2.08E-18 & 1.73E+01  \\
\hline
  -2 & 2.57E-04 & 1.49E-01 & 1.35E-02 & 3.33E+47 & 1.28E-15 & 8.42E-15 & 1.09E-14 & 1.49E+05 & 1.83E+05 & 4.62E-19 & 1.88E-18 & 1.45E+01  \\
\hline
  -1 & 2.97E-04 & 1.13E-01 & 4.72E-03 & 2.37E+47 & 4.64E-16 & 2.26E-15 & 2.88E-15 & 1.61E+05 & 1.97E+05 & 3.46E-19 & 1.63E-18 & 1.47E+01  \\
\hline
  0 & 3.56E-04 & 7.89E-02 & 4.25E-03 & 1.36E+47 & 4.87E-16 & 1.50E-15 & 1.85E-15 & 1.80E+05 & 2.20E+05 & 2.40E-19 & 1.32E-18 & 1.30E+01  \\
\hline
  1 & 3.84E-04 & 6.77E-02 & 5.66E-03 & 8.06E+46 & 7.14E-16 & 1.79E-15 & 2.16E-15 & 1.90E+05 & 2.32E+05 & 2.06E-19 & 1.21E-18 & 1.22E+01  \\
\hline
  2 & 5.15E-04 & 3.74E-02 & 1.08E-03 & 1.78E+46 & 2.16E-16 & 2.77E-16 & 2.72E-16 & 2.37E+05 & 2.91E+05 & 1.15E-19 & 8.95E-19 & 1.07E+01  \\
\hline
  3 & 6.64E-04 & 2.24E-02 & 3.21E-04 & 8.80E+45 & 1.03E-16 & 9.78E-17 & 7.17E-17 & 2.96E+05 & 3.63E+05 & 6.90E-20 & 7.03E-19 & 8.95E+00  \\
\hline
  4 & 8.12E-04 & 1.49E-02 & 1.10E-04 & 4.43E+45 & 5.21E-17 & 4.51E-17 & 2.73E-17 & 3.55E+05 & 4.34E+05 & 4.63E-20 & 5.64E-19 & 8.00E+00  \\
\hline
  5 & 9.60E-04 & 1.06E-02 & 4.29E-05 & 2.37E+45 & 2.82E-17 & 2.36E-17 & 1.31E-17 & 4.15E+05 & 5.08E+05 & 3.32E-20 & 4.66E-19 & 7.06E+00  \\
\hline
  6 & 1.11E-03 & 7.95E-03 & 1.84E-05 & 1.32E+45 & 1.60E-17 & 1.31E-17 & 7.00E-18 & 4.74E+05 & 5.81E+05 & 2.52E-20 & 3.84E-19 & 6.01E+00  \\
\hline
\end{tabular}
\end{center}
\caption{The frequency $\nu$ $({\rm Hz})$ and the respective amplitudes of horizontal surface displacement $\zeta_h(R_\odot)$, quadrupole moment $J$, the total kinetic energy of the mode $E$ (${\rm gm\cdot cm^2/sec^2}$), gravitational signals $S_m$ and velocity amplitudes $V_m$ (${\rm m/sec}$), for unit radial surface displacement $\zeta_r(R_\odot) = 1$ as a function of the radial order of the mode $n$ (negative and positive $n$ correspond to g-modes and p-modes respectively). The last three columns show the expected instrumental noise $B_i$ ($1/\sqrt{\rm Hz}$), binary confusion noise $B_b$ ($1/\sqrt{\rm Hz}$) and surface velocity noise $B_v$ (${\rm m/sec}/\sqrt{\rm Hz}$) respectively for the LISA mission at the corresponding frequencies.}
\label{table1}
\end{table}

The sun is rotating with a period $\sim 27$ days so a frame of reference fixed relative to the oscillating sun rotates relative to an inertial frame, producing a modulation of the time dependent gravitational field. The solar rotation axis is inclined to the ecliptic plane, and hence to the orbit plane of the gravitational detector, producing a further modulation of the signals from the oscillations and a very low frequency signal ($\Delta\nu \sim 4\times 10^{-7}~{\rm Hz}$) from the static quadrupole moment induced by the rotation. As the inclination is small (about $7^o$) we neglect these effects in the present analysis taking the rotation axis of the Sun perpendicular to the orbit plane of the detector.

Since the monopole gravitational field of the Sun is weak ($GM_\odot/rc^2 \sim 10^{-8}$ at 1 a.u.), and that of the time dependent quadrupole moments and any associated gravitational radiation even weaker, the metric of space-time in the neighbourhood of the Earth is adequately described by the weak field limit as
\bea
ds^2 = g_{ik} dx^i\,dx^k  = (\eta_{ik} + h_{ik}) dx^i\,dx^k
= \left(1+{2U\over c^2}\right) c^2 dt^2 -
\left(1-{2U\over c^2}\right)dx^\alpha dx^\alpha
+ h^{GW}_{\alpha\beta} dx^\alpha dx^\beta,
\label{7}
\ena
where $\eta_{ik} =  diag (1,-1,-1,-1)$ is the Minkowski metric, $|h_{ik}|\ll1$, indeces are raised and lowered using the Minkowski tensor, and a summation over repeated indices is implied. Roman indeces $i,k = 0,1,2,3$ whereas Greek indeces $\alpha,\beta = 1,2,3$. The $h_{ik}$ have a contribution from the time dependent quadrupole moments
\bea
h^N_{ik} = {2\,\delta U\over c^2} \delta_{ik} =
 -{G\,M_\odot R^2_\odot\over 3 c^2}\,\delta_{ik}\,{\cal D}^{\alpha\beta}
\nabla_{\alpha\beta}\left(1\over r\right)
=  -{G\,M_\odot R^2_\odot\over 3 c^2}\,\delta_{ik}\,
\nabla_{\alpha\beta}\left(1\over r\right)\,
\sum_{n=-\infty}^\infty\,\sum_{m=-2}^2 C_m\,J_{nm}{\cal I}_{m}^{\alpha\beta},
\label{8}
\ena
and a space-like contribution from the gravitational quadrupole radiation, which with time dependence $\propto e^{i\omega t}$ is
\bea
h^{GW}_{\alpha\beta} =
{2G\over 3 c^4 r} {d^2\over dt^2}\left(\tilde{\cal D}_{\alpha\beta}\right)
=-{2\omega^2 G\over 3 c^4 r} \tilde{\cal D}_{\alpha\beta}
=-{2\omega^2 G\,M_\odot R^2_\odot\over 3 c^4 r}\,\sum_{n=-\infty}^\infty\,
\sum_{m=-2}^2 C_m\,J_{nm}\delta_{\alpha\gamma}\,\delta_{\beta\epsilon}\,
\tilde{\cal I}_{m}^{\gamma\epsilon}.
\label{9}
\ena
In the above expression $\tilde{\cal D}_{\alpha\beta}$ is the retarded, transverse (i.e.~lying in the plane perpendicular to the radial direction of propagation of the waves) and trace-free part of $\cal{D}_{\alpha\beta}$. $\tilde{\cal I}_m^{\alpha\beta}$ are the retarded, transverse, trace-free tensors corresponding to the basis tensors ${\cal I}_m^{\alpha\beta}$.

The subdivision in expression (\ref{7}) of the metric field into the Newtonian potential and gravitational wave field at the border between the near and the wave zones is obviously a serious over simplification. Nevertheless, this subdivision picks out the essential physical aspects of the problem, and there are no obvious reasons for the exact analysis to yield qualitatively different answers. In a recent paper \cite{Kopeikin2006} the authors conducted a thorough analysis of light propagation through the intermediate zone between the wave and the near zones. In following papers we hope to analyze the problem of solar oscillations and their detectability by LISA using the exact analysis in \cite{Kopeikin2006}.

%%%%%%%%%%%%%%%%%%%%%%%%%%%%%%%%%%%%%%%%%%%%%%%%%%%%%%%%%%%%%%%%%%%%%%%%%%%%%%%%%%%%%%%%
%%%%%%%%%%%%%%%%%%%%%%%%%%%%%%%%%%%%%%%%%%%%%%%%%%%%%%%%%%%%%%%%%%%%%%%%%%%%%%%%%%%%%%%%

\section{Phase shift in a perturbed gravitational field\label{SectionIII}}

In order to analyze the response of a LISA type interferometer to solar oscillations we firstly need to study the relative phase shift for light traveling along the arms of the interferometer in the perturbed gravitational field (\ref{7}). In order to proceed, let us consider an electromagnetic wave of frequency $\omega_e$ and wave 4-vector $k_i$ propagating along the arm of a detector from an emmitter at coordinate position $x_A^i$ to a receiver at coordinate position $x_B^i$. The phase $\phi$ of the wave at $x_B^i$ is given by
\bea
\int_A^B k_i dx^i =  (k_{iB}x_B^i - k_{iA}x_A^i) - \int_A^B x^i dk_i
= k_{iA}(x_B^i - x_A^i) + \int_A^B (x_B^i-x^i) dk_i .
\label{10}
\ena
In Minkowski space the wave vector $k^i$ is constant $=k_A^i = \omega_e (1,n^\alpha)/c$ and the phase difference $\phi(B)-\phi(A)$ is simply $k_{iA}(x_B^i - x_A^i)$. The second term on the right hand side (\ref{10}) is identically zero in the Minkowski limit.

In the time-dependent space time $g_{ik}$ the first order perturbation in phase, due
both to the departure from Minkowski space time and the displacement of the receiver B
 (taking A as fixed) is given by
\bea
\delta \varphi = k_{iA} \delta(x^i_B-x^i_A) + \int_A^B (x^i_B - x^i) dk_i .
\label{11}
\ena

We take $A$ as the origin of coordinates so $x_A^i=0$, and take the unperturbed ray to be given by $x^i = \lambda k^i$, where $\lambda$ is an affine parameter varying from $0$ at $A$, to $\Lambda$ at $B$. Since $k^i = dx^i/d\lambda$, the null geodesic equation in this weak field approximation reduces to
\bea
{dk^i\over d\lambda} =
-\Gamma^i_{mn} k^m k^n \approx {1\over 2} h_{mn,i} k^m k^n.
\nonumber
%\label{12}
\ena
and the last term in ({\ref{11}) can be expressed as
\bea
\int _0^\Lambda (\Lambda - \lambda) k^i dk_i = {1\over 2}
\int _0^\Lambda (\Lambda - \lambda) h_{mn,i} k^m k^n k^i d\lambda=
{1\over 2}
\int _0^\Lambda (\Lambda - \lambda) {dh_{mn}\over d\lambda} k^m k^n d\lambda.
\nonumber
%\label{13}
\ena
Substituting this result into (\ref{11}), defining $l^i=(x^i_B-x^i_A),~ \delta l^i=\delta(x^i_B-x^i_A)$, and integrating by parts gives
\bea
\delta \varphi = k_{iA} \delta l^i + \left[{1\over 2}
(\Lambda-\lambda) h_{mn} k^m k^n\right]_0^\Lambda +
 {1\over 2} \int_0^\Lambda h_{mn} k^m k^n d\lambda
\label{14}.
\ena
$\delta l^i$ is given by the geodesic deviation equation
\bea
{D^2\over D\tau^2}(\delta l^i) =  R^i_{jkm} {dx^j\over d\tau}
{dx^k\over d\tau} l^k ,
\nonumber
%\label{14a}
\ena
which, in the slow motion approximation appropriate to the current analysis, reduces to
\bea
{d^2\over dt^2}(\delta l^i) =  R^i_{oko} l^k.
%\nonumber
\label{15}
\ena
Since in the weak field approximation
\bea
R^j_{oko} = \Gamma^j_{ok,o} - \Gamma^j_{oo,k} + \Gamma^j_{oi} \Gamma^i_{ok}
- \Gamma^j_{ki} \Gamma^i_{oo} \approx
{1\over 2}\eta^{ij}\left(h_{ki,oo} - h_{oi,ok} + h_{oo,ik} +h_{ok,io}\right),
%\nonumber
\label{16}
\ena
From (\ref{16}) and (\ref{15}) it follows that $R^o_{oko} = 0$ and $d^2 \delta l^o/dt^2 = 0$, respectively. Recalling that $h_{o\alpha}=0$, and $l^i = l_0^i + \delta l^i$ with $l_0^i = \left(1,n^{\alpha}\right)$ we obtain
\bea
{d^2\over dt^2}\left(\delta l^\alpha\right)
 = {1\over 2}\left[
{d^2\over dt^2}\left( h_\beta^\alpha\right) -
c^2 \eta^{\alpha\gamma} \nabla^2_{\gamma \beta} h_{oo}\right] l_0^\beta,
\label{17}
\ena
where $\nabla^2_{\gamma \beta} =\partial^2/\partial x^\gamma\partial x^\beta$. In view of the time dependence of $h_{ik} \propto e^{i\omega t}$, this equation can be integrated to give
\bea
\delta l^\alpha = {1\over 2} \left( h^\alpha_\beta - {c^2
\eta^{\alpha\gamma}\over \omega^2} \nabla^2_{\gamma \beta} h_{oo}\right)
l_0^\beta.
\label{18}
\ena
Now since $k^\alpha=\omega_e n^\alpha /c, ~\Lambda=cl/\omega_e$, where $\omega_e$ is the frequency of the electromagnetic wave, substitution into (\ref{14}) gives
\bea
\delta \varphi = {\omega_e l\over 2 c} \left( h_{\alpha\beta}  n^\alpha
n^\beta - {c^2
\over \omega^2} n^\alpha n^\beta \nabla^2_{\alpha \beta} h_{oo}\right)
- {\omega_e l\over 2 c} h_{mn} n^m n^n + {\omega_e\over 2}  \int_0^{l/c} h_{mn}
n^m n^n dt,
\label{19}
\ena
where we have introduced $l = n_\alpha l_0^\alpha$ as the unperturbed distance between $A$ and $B$.

We now write $\delta\varphi = \delta\varphi_N +\delta\varphi_{GW}$, where $\delta\varphi_N$ is the contributions from the Newtonian potential $\delta U$ and $\delta\varphi_{GW}$ is the contribution from gravitational waves. Using expression (\ref{7}) and expanding the $h_{mn}$ in (\ref{19}) by Taylor series yields
\bea
\begin{array}{l}
\delta\varphi_N =  - {\omega_e l\over 2 c} h_{oo} - {\omega_e l c\over 2
\omega^2}
n^\alpha n^\beta \nabla^2_{\alpha \beta} h_{oo}
+{1\over 2}{\omega_e } h_{oo} \int_0^{l/c} e^{i\omega t} dt \\
~~~~~~~~~+ {1\over 2}\omega_e n^\alpha \nabla_\alpha h_{oo}
 \int_0^{l/c} e^{i\omega t} c t dt
+{1\over 4} \omega_e n^\alpha n^\beta \nabla^2_{\alpha\beta} h_{oo}
 \int_0^{l/c} e^{i\omega t} c^2 t^2 dt + \dots,
\\ \\
\delta\varphi_{GW} = {1\over 2} \omega_e \int_0^{l/c} h^{GW}_{\alpha\beta}
n^\alpha n^\beta dt \approx {1\over 2} {\omega_e l\over c} n^\alpha n^\beta
h^{GW}_{\alpha\beta}.
\end{array}
\label{21}
\ena
Taking into account (\ref{8}) and (\ref{9}), retaining just the leading terms in powers of $\omega l/c \ll1$ and $l/r \ll1$ in (\ref{21}), the phase shift is given by
\bea
\delta\varphi = -{\omega_e l\over 2 c} n^\alpha n^\beta \left({c^2\over
\omega^2}\nabla^2_{\alpha\beta} h_{oo} - h^{GW}_{\alpha \beta}\right)
={\omega_e l\over c}{G\over 6 \omega^2} n^\alpha n^\beta
\left({\cal D}^{\gamma\epsilon} \nabla^4_{\alpha\beta\gamma\epsilon} \left(1\over r\right) +
{2\omega^4\over c^4 r} \tilde{\cal D}_{\alpha\beta}\right),
\label{22}
\ena
where $\nabla^4_{\alpha\beta\gamma\delta} = {\partial^4/\partial x^\alpha \partial x^\beta \partial x^\gamma \partial x^\epsilon}$.
Since the unperturbed phase shift is $\varphi = \omega_e l/c$, the fractional change in phase shift can be expressed in the form
\bea
\left({\delta\varphi\over\varphi}\right) = {G\over 2\omega^2 r^5}
\left(T_{\alpha\beta}{\cal D}^{\alpha\beta}  + {2\omega^4 r^4\over 3 c^4}
 N_{\alpha\beta}\tilde{\cal D}^{\alpha\beta}\right),
 \label{23}
\ena
where  $N_{\alpha\beta}= n_\alpha n_\beta$, and
\bea
T_{\alpha\beta} = {1\over 3}\,r^5 N^{\gamma\epsilon}
\nabla^4_{\alpha\beta\gamma\epsilon}\left(1\over r\right) =
2 n_\alpha n_\beta - 10\mu (n_\alpha \bar{n}_\beta+\bar{n}_\alpha n_\beta)
+ 5 (7 \mu^2 - 1) \bar{n}_\alpha \bar{n}_\beta,
\label{24}
\ena
and $\bar{n}^\alpha$ is the unit vector in the radial direction and $\mu = n^{\alpha} \bar{n}_\alpha$. The details of the derivation of $T_{\alpha\beta}$ are given in Appendix \ref{AppendixII}.

%%%%%%%%%%%%%%%%%%%%%%%%%%%%%%%%%%%%%%%%%%%%%%%%%%%%%%%%%%%%%%%%%%%%%%%%%%%%%%%%%%%%%%%%
%%%%%%%%%%%%%%%%%%%%%%%%%%%%%%%%%%%%%%%%%%%%%%%%%%%%%%%%%%%%%%%%%%%%%%%%%%%%%%%%%%%%%%%%

\section{Response of the interferometer \label{SectionIV}}

A laser interferometer detector such as LISA consists of 3 arms AB, AC, CB, in circular orbit around the Sun. The interferometer's response is given by the difference in the fractional change in phase shifts of the round trip signals along any two of the arms, for example ABA and ACA, as
\bea
S = \left(\delta\varphi\over\varphi\right)_{AB}- \left(\delta\varphi\over
\varphi\right)_{AC}=
{G\over 2\omega^2 r^5}
\left( \Delta T_{\alpha\beta}{\cal D}^{\alpha\beta}+ {2\omega^4 r^4\over 3 c^4}
 \Delta N_{\alpha\beta}\tilde{\cal D}^{\alpha\beta}\right)
\label{25}
\ena
where $\Delta$ indicates the difference between the arms AB and AC.

The response $S$ is a function of the angles $\chi, \iota, \psi, \upsilon$, the quadrupole moments $J_m$ and the frequencu $\omega$. $\chi$ is the angle between the radius vector to the detector and the instantaneous orientation of the reference axes corotating with the sun, $\iota$ the inclinations of the plane of the detector to the orbit plane, $\psi$ the angle between the arm AB and the direction of the orbit and $\upsilon$ the angle between the two arms of the detector.  In the LISA concept, $\iota=\upsilon = \pi/3$ and the detector rotates with the same period as it orbits the Sun. The angle $\psi$ decreases as  the position angle $\phi$ of the detector (on its orbit relative to a fixed inertial frame) increases, i.e.~$\psi = \psi_0 - \phi$ where $\psi_0$ is a constant equal to the orientation of the detector relative to the orbit plane at the arbitrary zero of the orbit angle $\phi$.

We now express the quadrupole moment tensor in terms of the trace-free basis tensors ${\cal I}_m^{\alpha\beta}$ (see (\ref{3})) and determine the response $S_m$ for both gravitational waves and the Newtonian signal for quadrupole modes of azimuthal order $m$ and frequency $\nu = \omega/2\pi$. Introducing the effective gravitational amplitudes 
\bea
h_{eff}=2S_m,
\nonumber
\ena
we have
\bea
h_{eff} = h_{eff}^{GW} + h_{eff}^{N},
\nonumber
\ena
where the Newtonian and gravitational wave contributions to the signal can be represented in the form
\bea
\begin{array}{l}
h_{eff}^{GW}={2\over 3} C_m\,J_m\,\left(R_\odot\over r\right)^5
\left(\nu_\odot\over\nu\right)^2 \,
\left(\nu\over\nu_r\right)^4\,f^{GW}_m,
 \\ \\
h_{eff}^N=\,C_m\,J_m\,\left(R_\odot\over r\right)^5
\left(\nu_\odot\over\nu\right)^2 \,f^N_m,
\end{array}
\label{26a}
\ena
where (see Appendix \ref{AppendixIII} for details) 
\bea
\begin{array}{l}
f^{GW}_m=\tilde{\cal I}_m^{\alpha\beta} \Delta N_{\alpha\beta},
 \\ \\
f^N_m= {\cal I}_m^{\alpha\beta} \Delta T_{\alpha\beta},
\end{array}
\label{26}
\ena
and
\bea
\nu_\odot \equiv {1\over 2\pi}\sqrt{G\,M_\odot\over R^3_\odot}
\approx 10^{-4} {\rm Hz},~~~\nu_r \equiv \left(c\over 2\pi r\right) \approx 3\times 10^{-4}{\rm Hz}.
\nonumber
\ena
Note that, as follows from (\ref{26a}), for $f_m^N$ and $f_m^{GW}$ of comparable order the gravitational wave contribution dominates for $\nu>\nu_r\sim 3\times 10^{-4}~{\rm Hz}$.

To determine the functions $f^N_m, f^{GW}_m$ consider the detector to be at the point $P$ on its orbit with orbit angle $\phi$ relative to a fixed inertial frame, and let $\chi$ be the angle between the radius vector to $P$ and the $x^1$ axis of the corotating system $x^\alpha$. We then define local transverse Cartesian coordinates $\xi^\gamma$ at $P$ such that $\xi^1$ is in the outward radial direction, $\xi^2$ in the direction of the orbit and $\xi^3$ perpendicular to the orbit plane. The basis tensors in the $\xi^\gamma$ coordinates are given by
\bea
{\sl I}_m^{\gamma\epsilon} = e^\gamma_\alpha e^\epsilon_\beta {\cal I}_m^{\alpha\beta},~~
\nonumber
\ena
where
\bea
e^1 = (\cos\chi, \sin\chi, 0),~
e^2 = (-\sin\chi, \cos\chi, 0),~e^3 = (0, 0, 1),
\nonumber
%\label{27}
\ena
are the unit vectors along the $\xi^\gamma$ axes in the $x^\alpha$ coordinate system. The transverse trace-free tensors $\tilde {\sl I}^{\gamma\epsilon}$ are then given by (see Appendix \ref{AppendixIII})
\bea
\tilde {\sl I}_m^{23} = \tilde {\sl I}_m^{32} = {\sl I}_m^{23},~~\tilde {\sl I}_m^{22}
 = -\tilde {\sl I}_m^{33} =
{1\over 2}\,({\sl I}_m^{22}-{\sl I}_m^{33}),~~~
\tilde {\sl I}_m^{1 \gamma} = 0.
\nonumber
%\label{28}
\ena
In the $\xi^\gamma$ coordinates, in light of the definitions of angles $\iota$, $\psi$ and $\upsilon$, the unit vectors $n_B^\gamma, n_C^\gamma$ are given by
\bea
n_B^\gamma = (-\sin\psi\cos\iota,
\cos\psi,\sin\psi\sin\iota),~~n_C^\gamma = (-\sin\psi'\cos\iota,
\cos\psi',\sin\psi'\sin\iota),~\psi'=\psi+\upsilon.
\nonumber
%\label{29}
\ena
In light of the definition the coordinate system $\xi^\gamma$, the unit radial vector $\bar{n}^{\gamma}$ is given by
\bea
\bar{n}^\gamma = (1,0,0),
\nonumber
\ena
After some laborious algebra we determine the tensors $\Delta N_{\gamma\epsilon}, \Delta T_{\gamma\epsilon}$ in the $\xi^\gamma$ coordinates, and hence the functions $f^N_m, f^{GW}_m$ defined in (\ref{26}). The solutions for arbitrary $\iota,\upsilon,\psi,\chi$  are given in Appendix \ref{AppendixIII}.

For LISA, setting $\iota=\upsilon = \pi/3$ and $\psi_0=0$ (i.e.~$\psi=-\phi$), we obtain
\bea
\begin{array}{l}
f^{N}_0 ={3\sqrt{3}\over 8}\,\sin(2\psi+\pi/3),~~f^{GW}_0 =
{21\sqrt{3}\over 16}\,\sin(2\psi+\pi/3),
\\ \\
f^N_1 = -3\,\left[\,2\,\cos\chi\,\sin(2\psi+\pi/3)
-\sin\chi\,\cos(2\psi+\pi/3)\,\right],~~f^{GW}_1 =
{3\over 2}\,\sin\chi\,\cos(2\psi+\pi/3), 
\\ \\
f^N_2 = -{\sqrt{3}\over 2}\,\left[{25\over 4}\,\cos{2\chi}\,
\sin(2\psi+\pi/3) - 8\,\sin{2\chi}\,\cos(2\psi+\pi/3)\,\right],~~
f^{GW}_2 = -{7\sqrt{3}\over 16}\,\cos{2\chi}\,\sin(2\psi+\pi/3).
\label{30} 
\end{array}
\ena
The values for $m=-1,-2$ are obtained from those for $m=1,2$ by replacing $\chi$ by $\chi-\pi/2m$.

Relative to a fixed inertial frame at time $t$ the detector is at orbit angle $\phi = \Omega_d t$ and the corotating $x^1$ axis is at an angle $\Omega_\odot t$, where $\Omega_d$ is the angular velocity of the detector around the sun and $\Omega_\odot$ the angular velocity of the Sun relative to this inertial frame.  The angle between the radius vector from the sun to the detector and the rotating $x^1$ axis is therefore $\chi =\Omega_s t$ where $\Omega_s= 2\pi/P_s$ where $P_s\sim 26.75$ days is the synodic period of solar rotation, that is the period relative to a reference frame orbiting the sun at 1 a.u. Thus, substituting $\psi=-\phi$ and $\chi=\left(\Omega_s/\Omega_d\right)\phi$ into (\ref{30}) gives $f^N_m(\phi)$ and $f^{GW}_m(\phi)$ as a function of the position of the detector on its orbit determined by the orbit phase angle $\phi$. In Figure \ref{figure2a2c}, for illustration, we show the $f^N_m(\phi)$ and the $f^{GW}_m(\phi)$ modes. Since $m=0$ correspond to axially symmetric modes, the functions $f_0(\phi)$ are independent of the rotation of the Sun. The other modes ($m=\pm1,\pm2$) display the modulation of the signal due to solar rotation. Combining the contributions from $f^N$ and $f^{GW}$ gives
\bea
\begin{array}{l}
S_0(\nu,\phi) = {3\sqrt{3}\over 16}\,C_0\,J_0\,{R^5_\odot\over r^5}\,
{\nu^2_\odot\over\nu^2}\,\left(1+{7\over 3}\,{\nu^4\over\nu_r^4}\right)\,\sin(2\psi+\pi/3), \\ \\
S_1(\nu,\phi) = {3\over 2}\,C_1\,J_1\,{R^5_\odot\over r^5}\, {\nu^2_\odot\over\nu^2}\,\left[\,
\left(1 + {1\over 3}\,{\nu^4\over\nu_r^4}\right) \sin\chi\,\cos(2\psi+\pi/3)
-2\,\cos\chi\,\sin(2\psi+\pi/3)\right],
\\ \\
S_2(\nu,\phi) = 2\sqrt{3} C_2\,J_2\,{R^5_\odot\over r^5}\,{\nu^2_\odot\over\nu^2}\,
\left[\sin{2\chi}\,\cos(2\psi+\pi/3)
 -{25\over 32}\,\left(1 + {7\over 75}\,{\nu^4\over\nu^4_r}\right)\cos{2\chi}\, \sin(2\psi+\pi/3)\right],
\end{array}
\label{31}
\ena
where
\bea
\psi = -\phi,~~\chi = \phi P_d/P_\odot \approx 13.65\,\phi,~~C_{0}
 = -\sqrt{5\over 4\pi},~~C_{m} = \sqrt{15\over 4\pi},
~~~m=\pm1,\pm2.
\nonumber
%\label{32}
\ena
Again, in a similar fashion to (\ref{30}), values for $m=-1,-2$ are obtained from those for $m=1,2$ by replacing $\chi$ by $\chi-\pi/2m$. The expressions in (\ref{31}) contain both the contributions, from the Newtonian potential and gravitational waves (terms proportional to $\left(\nu/\nu_r\right)^4$). As anticipated above, the gravitational wave contribution dominates when $\nu>\nu_r \sim 3\times 10^{-4}~{\rm Hz}$.

\begin{figure}
\begin{center}
\includegraphics[width=8cm]{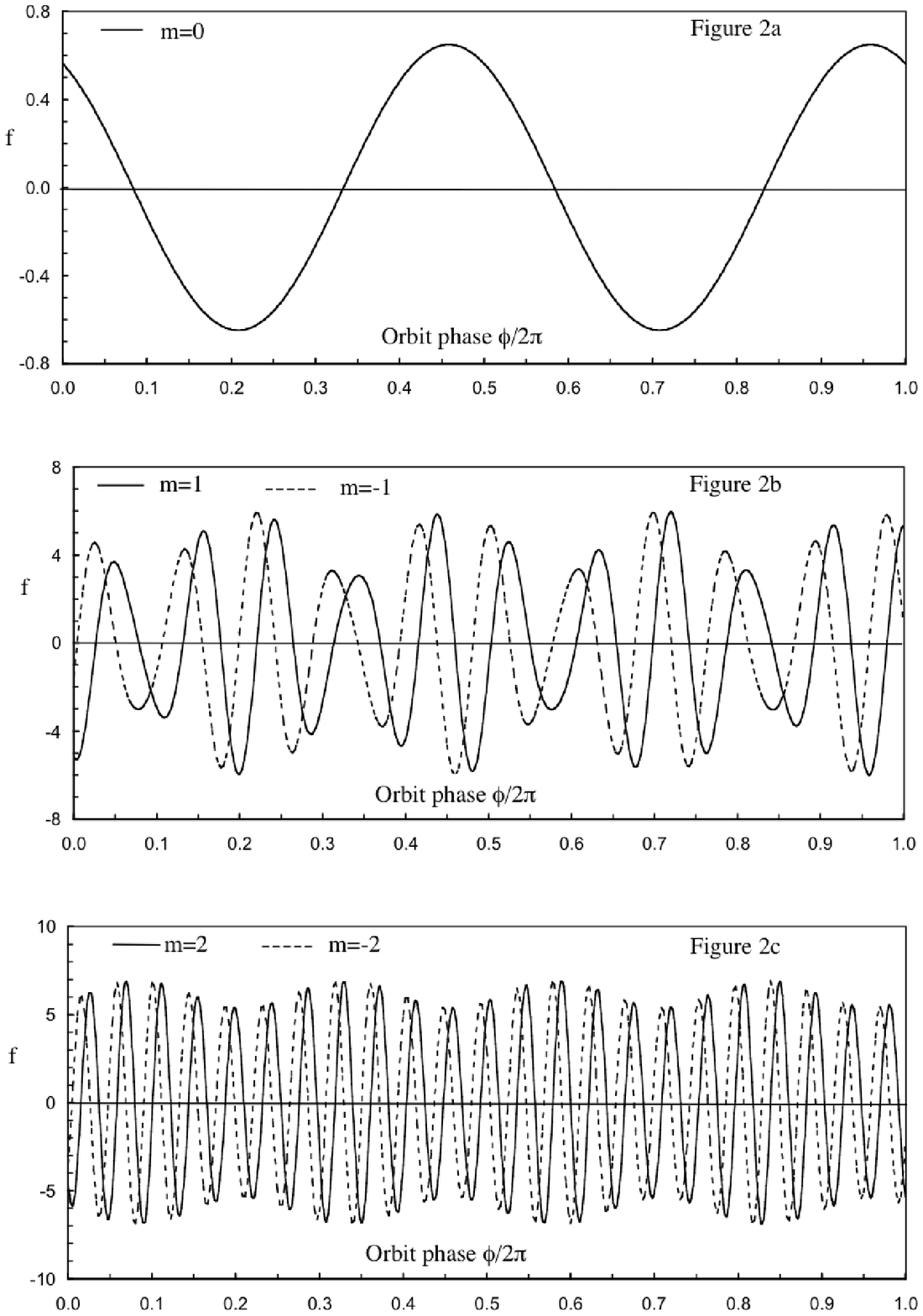} \quad
\includegraphics[width=8cm]{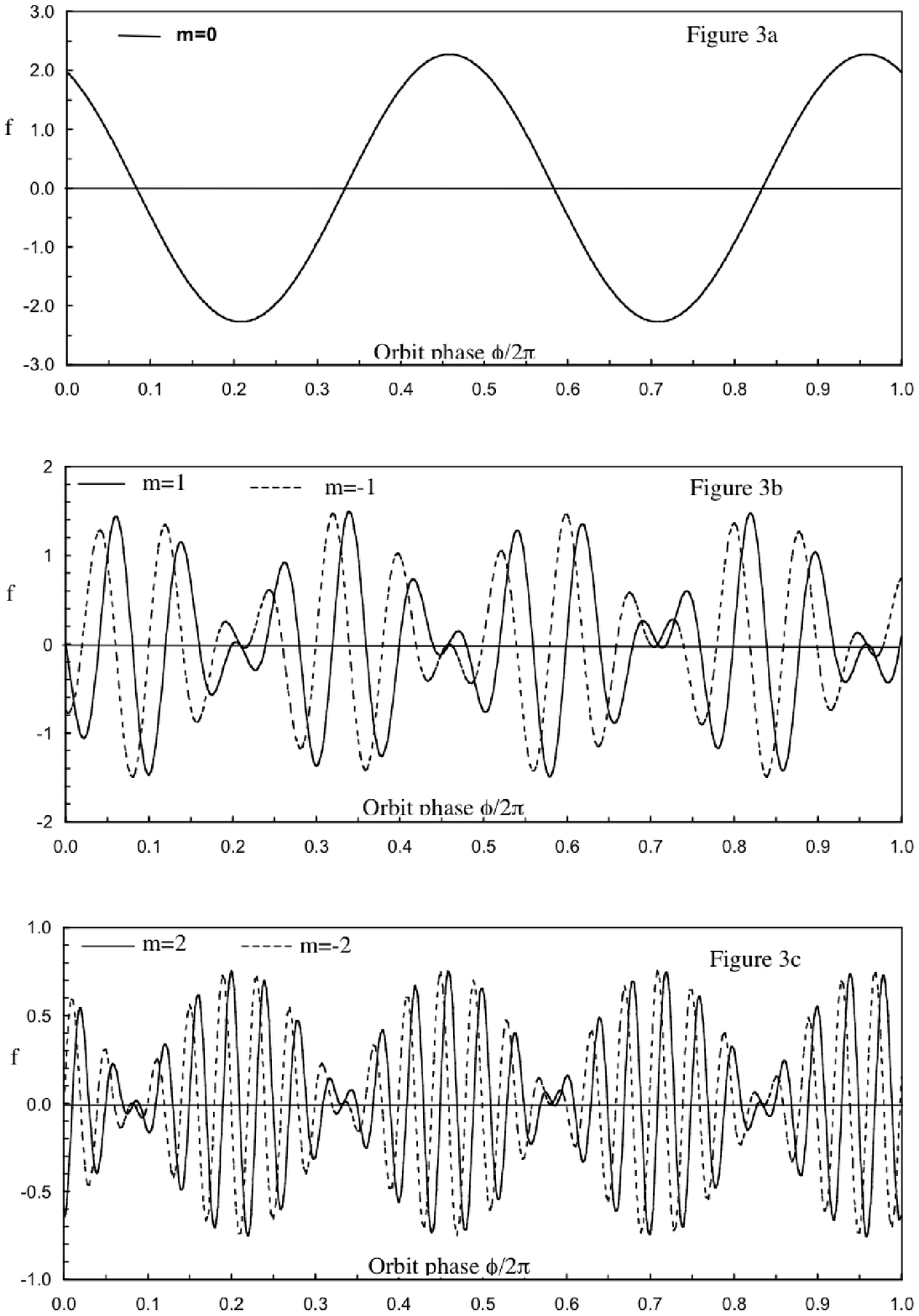}
\end{center}
\caption{Signal functions $f^{N}_m(\phi)$ (left panel) and $f^{GW}_m(\phi)$ (right panel) for gravitational waves from oscillations as a function of phase angle for modes $\ell = 2$, $m=0,\pm1,\pm2$.}\label{figure2a2c}
\end{figure}

The amplitude of the signal of given $(m,\nu)$ varies with the rotation of the sun around its axis and over the coarse of the year. For an observation time $T \sim 1~{\rm year}$ the amplitude is given by the root mean square averaged over $T$. The resulting amplitudes are given in Table \ref{table1}  (columns $5,~6 ~{\rm and}~7$ for $m = 0,~1~{\rm and}~2$ respectively) for values of the quadrupole moments given in column $4$. These signals scale linearly with the value of the quadrupole moments.

%%%%%%%%%%%%%%%%%%%%%%%%%%%%%%%%%%%%%%%%%%%%%%%%%%%%%%%%%%%%%%%%%%%%%%%%%%%%%%%%%%%%%%%%
%%%%%%%%%%%%%%%%%%%%%%%%%%%%%%%%%%%%%%%%%%%%%%%%%%%%%%%%%%%%%%%%%%%%%%%%%%%%%%%%%%%%%%%%

\section{Detectability of Gravitational Signals from the Sun and comparison with Velocity Experiments\label{SectionV}}

The expected sensitivity of the LISA experiment for a single 2-arm detector has been evaluated in \cite{LISA_PrePhaseA}. At low frequencies the dominant contribution is from the acceleration noise $\propto 1/\omega^2$, at intermediate frequencies from shot noise, and at high frequencies the sensitivity declines when the path length along the detector becomes comparable to and greater than the wavelength of the gravitational wave. In the frequency range of interest this instrumental noise level is given by
\bea
h_I(\nu) \approx \left(1\over \sin i\right)
\left(2\times 10^{-41} + 1\times 10^{-51}{1\over\nu^4}\right)^{1/2}
 /{\sqrt{Hz}},
\label{33}
\ena
where $i=\pi/3$ is the angle between the arms of the interferometer. In addition to the instrumental noise there is also a "confusion noise", $h_B(\nu)$, due to the integrated gravitatonal wave contribution from galactic and extragalactic binary systems. The magnitude of this "confusion noise", $h_B(\nu)$, is uncertain but has been estimated in \cite{Bender1997} and their estimate is given in column 12 of Table 1 and shown in Figure \ref{figure2}.

Thus, for LISA, the threshold of detectibility $h_{TD}$ at a signal to noise level $S/N$, when direction to the source is known, is given by
\bea
h_{TD} \approx(h_I+h_B)\left(1\over \sqrt{T}\right) \left(S\over N\right),
\label{33a}
\ena
where $T$ is the time of observation (in secs).

\begin{figure}
\begin{center}
\includegraphics[width=12cm]{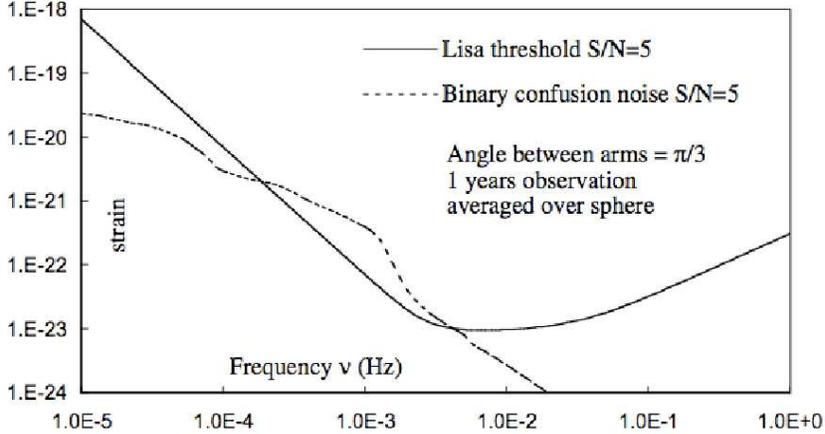}
\end{center}
\caption{Limits on detectable signal from LISA with signal to noise S/N=5.
(a) Instrumental noise, (b) binary confusion noise (Bender and Hils 1997).}\label{figure2}
\end{figure}

Let us now turn to the sensitivity of velocity experiments. The solar oscillations of low degree are obtained by taking the power spectrum of a time series of measurements of the Doppler shift (or velocity) of a given spectral line in the integrated light from the sun. The frequencies are the peaks in this power spectrum.  The velocity amplitude of a given quadrupole oscillation is therefore given by integrating the component of the surface velocity in the direction of the observer over the visible solar disc,
\bea
V = {{\int \tilde{k}^\alpha v_\alpha\,f(\mu)\,\mu\,dS}\over
{\int f(\mu)\,\mu\,dS}},
\label{35}
\ena
where $\tilde{k}^\alpha$ is the unit vector in the direction of the observer, $\mu = \tilde{k}^\alpha \hat r_\alpha$ is the cosine of the angle between the unit radius vector $\hat r_\alpha$ and the direction to the observer, and $f(\mu)$ the appropriate solar limb darkening function which incorporates the angular dependence of the intensity of radiation at the solar surface (see for example \cite{Allen1973}). We here take $f(\mu) = a + b\mu$ with $a=0.55, b=0.45$, which is a reasonable approximation for the whole disc velocity measurements by the GOLF \cite{Turck-Chieze1998} and BiSON \cite{Chaplin1998} experiments.

In a system of spherical coordinates $(r,\,\theta,\,\phi')$ corotating with the sun, the velocity on the solar surface due to an oscillation mode is
\bea
v_\alpha = (v_r, v_\theta, v_\phi) = \omega\,R_\odot\left(\zeta_r,\,\zeta_h
{\partial\over\partial\theta},\,
\zeta_h {1\over\sin\theta} {\partial\over\partial\phi'}\right)\,
S_{2 m}(\theta, \phi'),
\nonumber
%\label{36}
\ena
where $\zeta_r, \zeta_h$ are the dimensionless radial and horizontal displacement eigenfunction at the solar surface $R=R_\odot$. Let the corotating axes be at an angle $\phi_0$ to the direction to the observer and take $\phi = \phi'-\phi_0$ then
\bea
\tilde{k}^\alpha v_\alpha = v_r \sin\theta\cos\phi + v_\theta \cos\theta\cos\phi -
v_\phi \sin\phi,~~~~\mu = \sin\theta\cos\phi.
\nonumber
%\label{37}
\ena
Evaluating the integral (\ref{35}) gives the velocity amplitudes
\bea 
V_0 = -\omega\,R_\odot\,\sqrt{5\over 4\pi}\left({(16a+15b)\zeta_r +
6(8a+5b)\zeta_h\over 40(3a+2b)}\right),~~
V_{\pm 2} = \sqrt{3}V_0\left(\begin{array}{c}\cos 2\phi_0 \\ \sin 2\phi_0 \end{array}\right),
\label{38}
\ena
For $m = \pm 1, S_{2m} \propto \cos\theta$ is antisymmetric about $\theta = \pi/2$ and the integrals (\ref{35}) for $V_{\pm 1}$ are identically zero. The $m=\pm 1$ modes are not detectable in integrated velocity in the equatorial plane.

The surface horizontal displacement $\zeta_h$ is known in terms of the surface value of the radial displacement eigenfunction $\zeta_r$ from integration of the eigenvalue equations for the oscillation (column 3 in Table \ref{table1}). Since the normalised value of the quadrupole moment is $J_2\ll1$ for all modes (column 4), it follows from the equations governing the oscillations that $\zeta_h \approx \zeta_r\,g/\omega^2 \,R_\odot$ (see \cite{Unno1989}). The velocity amplitudes with $\zeta_r = 1$ are listed in columns 9 and 10. The background noise in velocity experiments is predominantly from velocity fields on the solar surface (active regions, granulation, meso-granulation, supergranulation) and the cumulative effect of these motions has been estimated in \cite{Harvey1995}. However in contrast to the gravitational case this background noise has been measured in a number of helioseismology experiments and in Figure \ref{figure3} we show the background velocity noise determined by the GOLF experiment on SOHO \cite{Gabriel1997,Turck-Chieze1998}. This is in reasonable agreement with the value obtained by the ground based whole disc networks BiSON \cite{Elseworth1994}, that obtained $900~{\rm cm/sec}/\sqrt{{\rm Hz}}$ at $\nu=5\times 10^{-4}~{\rm Hz}$ and IRIS experiment \cite{Fossat1996} which obtained  $2\times 10^3~{\rm cm/sec}/\sqrt{{\rm Hz}}$ at $\nu=10^{-4}~{\rm Hz}$. These values are a factor $\sim 5$ below the model predictions of \cite{Harvey1995}. In the region of interest this background noise $B_v(\nu)$ can be approximated by
\bea
B_v(\nu) \approx 2\times 10^3\,\left(10^{-4}\over \nu\right)^{1/2}\,~\frac{{\rm cm/sec}}{\sqrt{{\rm Hz}}}.
\label{39}
\ena

\begin{figure}
\begin{center}
\includegraphics[width=12cm]{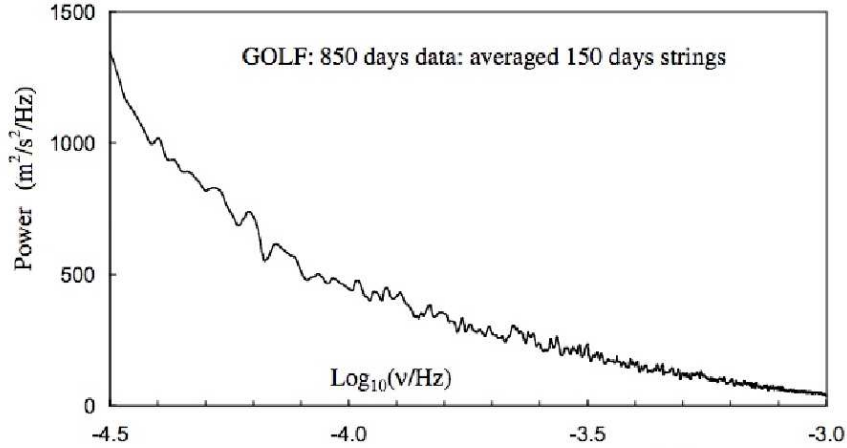}
\end{center}
\caption{Solar background noise in velocity (${\rm m^2/sec^2/Hz}$) as determined from the GOLF experiment on SOHO.}\label{figure3}
\end{figure}

We now make the following comparison. We assume a frequency resolution in the power series analysis of $3\times 10^{-8}~{\rm Hz}$ (corresponding to 1 year's observation time) and the value of a velocity signal from an oscillation mode that has $S/N=3$ using the above background noise estimate (\ref{39}).  This gives a value of the dimensionless radial eigenfunction $\zeta_r$ through (\ref{38}) for the velocity amplitude for each frequency and $m=0,2$. This then determines the amplitude of the quadrupole moment $J$ and the gravitational signal strength $S$ obtained by simply scaling $S_m$ in Table \ref{table1} by the value of $\zeta_r$. We then compare this gravitational signal strength with the background noise for the LISA experiment computed from (\ref{33}) under the same assumptions. The results are shown in Figure \ref{figure4}.  We see that for frequencies $\nu < 2\times 10^{-4}~{\rm Hz}$ the signal from the m=2 modes are stronger in the gravitational experiment than in the velocity experiment. In Figure \ref{figure4} we also show the $S/N$ for the gravitational experiment when binary confusion noise is included. Note that since the averaged velocity for the $m=\pm 1$ modes is zero there is no limit on the gravitational signal set by the helioseismology experiments.

This comparison between the 2 experiments is of course independent of the assumed $S/N$ in the velocity experiment, what is being compared is the detectability of an oscillation mode by the two techniques and is essentially the ratio of the $S/N$ for the gravitational experiment to $S/N$ for the velocity experiment.

We can reverse the comparison and ask what would be the velocity amplitude of modes that are at the margin of detectibility with the gravitational detector, and how does this compare with the solar background noise? This is done in Figure \ref{figure5}, here we see that the velocity amplitude at low frequencies for the $m=2$ mode is well below the noise level of velocity experiments for a frequency resolution of $30$ nHz. For high frequencies $\nu \gtrsim 3\times 10^{-4} ~{\rm Hz}$ the gravitational signal is predominantly determined by the gravitational wave component. For comparison, the figure also shows values for the velocity amplitude obtained from \cite{Cutler1996}. These values were calculated using expression (3)  and the data in Table 1 in \cite{Cutler1996}, note that these values are averaged over $m$-modes. As can be seen, the estimates from \cite{Cutler1996} are in a reasonable agreement (taking into account the averaging over $m$-modes) with our results at lower frequencies $\nu \lesssim 3\times 10^{-4} ~{\rm Hz}$. Let us compare the results for frequencies $\nu \gtrsim 3\times 10^{-4} ~{\rm Hz}$. As was mentioned above, at these frequencies the major contribution to the signal comes from the gravitational waves. This gravitational wave contribution was not analyzed in \cite{Cutler1996}. For this reason, as can be seen on Figure \ref{figure5}, at higher frequencies \cite{Cutler1996} give higher values for the detectability threshold in comparison with present results. In other words, the incorporation of the gravitational wave contribution makes detection of solar oscillations by LISA more feasible.

\begin{figure}
\begin{center}
\includegraphics[width=12cm]{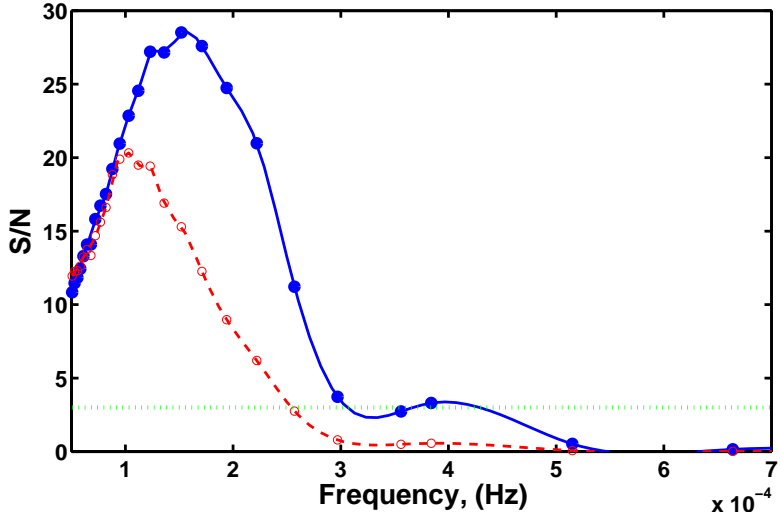}
\end{center}
\caption{Comparison of signal to noise ($S/N$) in LISA experiment for $S/N=3$ in solar velocity experiments, for oscillation modes with $\ell=2, m=2$. Solid line shows the $S/N$ when only the instrumental noise is included, while the dashed line shows the $S/N$ when both the instrumental noise and the noise from binary systems is included.}\label{figure4}
\end{figure}

\begin{figure}
\begin{center}
\includegraphics[width=12cm]{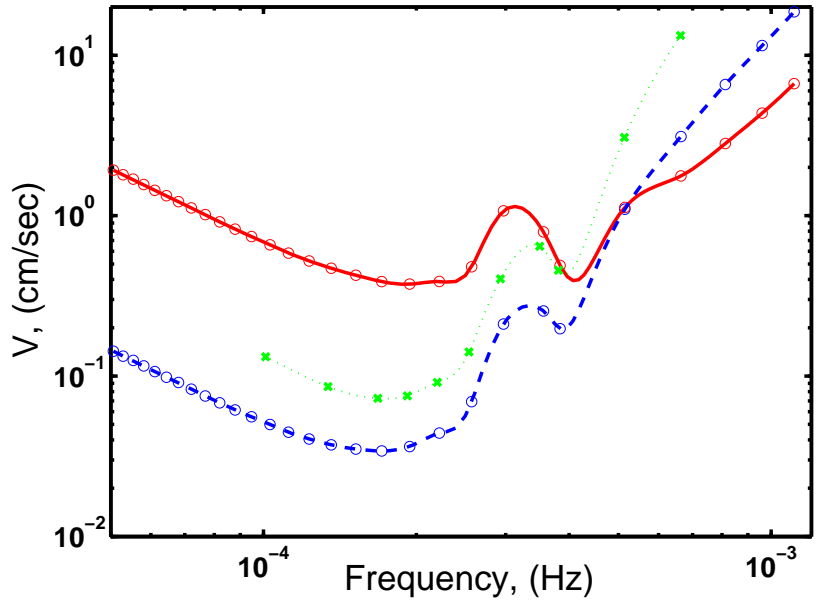}
\end{center}
\caption{Minimum surface velocity amplitude for signal to be detectable by LISA with $S/N=1$ for modes with $m=0$ (solid line) and $m=2$ (dashed line). The thin dotted line show the velocity amplitudes evaluated from \cite{Cutler1996}.}\label{figure5}
\end{figure}

In the above considerations we have assumed that the signal is monochromatic, or rather that the line is narrower than the frequency resolution.  This is the prediction from extrapolating the p-modes observed by helioseismology at higher frequencies. Since the rotational splitting of the modes is known, signal enhancement techniques can be used,  superposing power in frequency bins separated by $m\Omega_\odot$, as is being done in the search for g-modes in helioseismology by the Phoebus group \cite{Frohlich1998}. If the modes have a line width in excess of the frequency resolution the detectibility in both velocity and gravity is correspondingly reduced but the ratio of $S/N$ for the two experiments remains the same.

%%%%%%%%%%%%%%%%%%%%%%%%%%%%%%%%%%%%%%%%%%%%%%%%%%%%%%%%%%%%%%%%%%%%%%%%%%%%%%%%%%%%%%%%
%%%%%%%%%%%%%%%%%%%%%%%%%%%%%%%%%%%%%%%%%%%%%%%%%%%%%%%%%%%%%%%%%%%%%%%%%%%%%%%%%%%%%%%%

\section{Discussion and Conclusions\label{SectionVII}}

In this work we have analyzed the prospects of detecting the gravitational signal from solar oscillations in a LISA type interferometer, and compared them with capabilities of velocity experiments. At low frequencies, $\nu \lesssim 3\times 10^{-4}~{\rm Hz}$, the quadrupole oscillations of the Sun might be detected in LISA through their contribution to the time varying Newtonian (near zone) gravitational potential. At higher frequencies, $\nu>3\times 10^{-4}~{\rm Hz}$, LISA might observe a gravitational wave (far zone) signal from the solar oscillations. For frequencies $\nu \lesssim 2\times 10^{-4}~{\rm Hz}$ the signal will have a higher $S/N$ in a LISA type space interferometer than in helioseismology measurements, while for higher frequencies the signal would be more readily observable through helioseismology measurements.

Although, as was mentioned at the end of Section \ref{SectionII}, our analysis was based on the subdivision of the gravitational field into Newtonian potential and gravitational wave field at the border between the near and the wave zones, there is no reason to believe that an exact analysis will give qualitatively different results. In particular, the characteristic features in signal to noise ratio curve on Figure \ref{figure4} should remain, although there might be a slight change in the hight of the curve around frequencies $\nu\sim3\times10^{-4} ~{\rm Hz}$ corresponding to the intermediate zone. In a following paper we hope to analyze this question in more detail. 

The low frequency solar oscillation modes, considered in this paper, are more sensitive to the structure of the solar core than the higher frequency p-modes. Helioseismology experiments have so far been able to measure only the higher frequency p-modes. Hence, if the low frequency modes are detected by LISA, they would advance our understanding of the structure of this central core. This is because, the helioseismology experiments measure the surface velocity which is dependent on the detailed structure of the outer layers of the sun, and this has to be subtraced off to give diagnostic information on the solar interior. On the other hand, the interferometer experiments measure the actual quadrupole moments which are determined by the oscillations in the deep high density solar interior and therefore encode information about the solar interior which is independent of the structure and physics of the outer layers.

The predicted amplitudes of these low $n$ modes are still uncertain. The $p$-modes are thought to be stochastically excited by the convective motions, and if this is the excitation mechanism of the g-modes the predicted amplitudes of the low frequency modes are  very small.  But as was first pointed out in \cite{Dilke1972,Christensen-Dalsgaard1974}, the steep gradient of $^3$He in the inner half of the sun could also provide an excitation mechanism for these modes which may be damped by parametric resonance with other modes or by mild turbulent diffusion, but still be of sufficient amplitude to give rise to a detectable gravitational signal.

We note also that the frequency of $10^{-4}~{\rm Hz}$ corresponds to a period of the order of $160$ minutes.  There have been repeated suggestions that such a signal has been seen in helioseismology experiments, many but not all of the claims being withdrawn following more detailed analysis (see \cite{Kotov1997}). An analysis of the GOLF data \cite{Palle1998} placed an upper limit on the velocity amplitude of such a mode at about $1~{\rm cm/sec}$, an estimate compatible with searches by the Phoebus group \cite{Frohlich1998}. We note that for such a 160 minute mode an $S/N=1$ in the helioseismic velocity experiments would correspond to an $S/N=20$ in a LISA type experiment.

The search for g-modes in the helioseismology experiments is ongoing and hopefully will result in the confirmed detection of some modes prior to the launch of LISA. Were such modes to have been identified, they would provide a valuable known signal for the space interferometer which could then be used to calibrate and test the experiment, giving more credence to the interpretation of signals from more distant astrophysical sources.

%%%%%%%%%%%%%%%%%%%%%%%%%%%%%%%%%%%%%%%%%%%%%%%%%%%%%%%%%%%%%%%%%%%%%%%%%%%%%%%%%%%%%%%%
%%%%%%%%%%%%%%%%%%%%%%%%%%%%%%%%%%%%%%%%%%%%%%%%%%%%%%%%%%%%%%%%%%%%%%%%%%%%%%%%%%%%%%%%

\section*{Acknowledgements}
The authors are grateful to S. V. Vorontsov for valuable discussions on helioseismology. The authors would also like to thank S.~M.~Kopeikin, B.~.S.~Sathyaprakash and K.~S.~Thorne for valuable discussions and fruitful comments. 
We wish to thank Giacomo Giampieri, who contributed to the early work on this paper.

%%%%%%%%%%%%%%%%%%%%%%%%%%%%%%%%%%%%%%%%%%%%%%%%%%%%%%%%%%%%%%%%%%%%%%%%%%%%%%%%%%%%%%%%
%%%%%%%%%%%%%%%%%%%%%%%%%%%%%%%%%%%%%%%%%%%%%%%%%%%%%%%%%%%%%%%%%%%%%%%%%%%%%%%%%%%%%%%%

\appendix

\section{The quadrupole basis tensors ${\cal I}^{\alpha\beta}$ and surface harmonics $S_{2 m}$\label{AppendixI}}

The quadrupole gravitational potential can be expressed in the two equivalent forms (see (\ref{2}))
\bea
G\,M_\odot\,R_\odot^2 \sum_{m=-2}^2~\sum_{n=-\infty}^\infty
{J_{nm}\over r^3} S_{2m}(\theta, \phi) =
G\,{1\over 6}\,{\cal D}^{\alpha\beta}\, \nabla_{\alpha\beta}\left(1\over r\right),
\label{A1}
\ena
where $(r,\theta,\phi)$ are spherical polar coordinates, $x^\alpha = (x,y,z)$ Cartesian coordinates with $\theta =0$ the $z$ axis and $\phi = 0$, $\theta=\pi/2$ the $x$ axis,  and $\nabla_{\alpha\beta}=\partial^2 /\partial x^\alpha\partial x^\beta$.  $J_{nm}$ are the (dimensionless) quadrupole moments, ${\cal D}^{\alpha\beta}$ the quadrupole moment tensor and $S_{2m}$ are the real surface harmonics normalised to unity over the sphere
\bea
S_{2,0} = \sqrt{5\over 4\pi}\,{3\cos^2\theta-1\over 2},~~
S_{2,\pm 1} = \sqrt{15\over 16\pi}\,\sin 2\theta\,
\left(\begin{array}{c} \cos\phi \\ \sin\phi \end{array}\right),~~
S_{2,\pm 2} = \sqrt{15\over 16\pi}\,\sin^2\theta\, \left(\begin{array}{c} \cos 2\phi \\  \sin 2\phi \end{array}\right).
\nonumber
%\label{A2}
\ena
We define a set of 5 independent trace free tensors ${\cal I}_m^{\alpha\beta}$ as
\bea
{\cal I}_0= \left(\begin{array}{ccc} 1&0&0\\ 0&1&0\\ 0&0&-2 \end{array}\right),~
{\cal I}_1= \left(\begin{array}{ccc}  0&0&1\\ 0&0&0\\ 1&0&0 \end{array}\right),~
{\cal I}_{-1}=\left(\begin{array}{ccc} 0&0&0\\ 0&0&1\\ 0&1&0  \end{array}\right),~
{\cal I}_2=\left(\begin{array}{ccc} 1&0&0\\ 0&-1&0\\ 0&0&0  \end{array}\right),~
{\cal I}_{-2}= \left(\begin{array}{ccc}  0&1&0\\ 1&0&0\\ 0&0&0 \end{array}\right).
\nonumber
%\label{A3}
\ena
Taking into account
\bea
\nabla_{\alpha\beta}\left(1\over r\right) ={\partial^2\over\partial x^\alpha\partial x^\beta}
\left(1\over r\right) = \left(3 x_\alpha x_\beta - r^2 \delta_{\alpha\beta}\over r^5 \right),
\nonumber
%\label{A4}
\ena
where $\delta_{\alpha\beta}$ is the Kronecker delta, and
\bea
x^1 = x = r\sin\theta\cos\phi,~~x^2 = y = r\sin\theta\sin\phi,~~x^3 = z = r\cos\theta,
\nonumber
%\label{A5}
\ena
the ${\cal I}_m^{\alpha\beta}$ satisfy the relations
\bea
\label{A6}
\begin{array}{l}
{\cal I}_0^{\alpha\beta}\,\nabla_{\alpha\beta}\left(1\over r\right) = {3\over r^5}\,(x^2+y^2-2 z^2)
 = {6\over r^3}\,{1 - 3 \cos^2\theta\over 2} = -{6\over r^3}\,\sqrt{4\pi\over 5}\,S_{2,0},
\\ \\
{\cal I}_1^{\alpha\beta}\,\nabla_{\alpha\beta}\left(1\over r\right) = {3\over r^5}\,2xz = {6\over r^3}\,\sin\theta\cos\theta\cos\phi = {3\over r^3}\,\sqrt{16\pi\over 15}\,S_{2,1},
\\ \\
{\cal I}_{-1}^{\alpha\beta}\,\nabla_{\alpha\beta}\left(1\over r\right) = {3\over r^5}\,2yz = {6\over r^3}\,\sin\theta\cos\theta\sin\phi =
{3\over r^3}\,\sqrt{16\pi\over 15}\,S_{2,-1},
\\ \\
{\cal I}_2^{\alpha\beta}\,\nabla_{\alpha\beta}\left(1\over r\right) = {3\over r^5}\,(x^2-y^2) =  {3\over r^3}\,\sin^2\theta\cos 2\phi
={3\over r^3}\,\sqrt{16\pi\over 15}\,S_{2,2},
\\ \\
{\cal I}_{-2}^{\alpha\beta}\,\nabla_{\alpha\beta}\left(1\over r\right) = {3\over r^5}\,2xy =  {3\over r^3}\,\sin^2\theta\sin 2\phi
={3\over r^3}\,\sqrt{16\pi\over 15}\,S_{2,-2}.
\end{array}
\ena
We now expand the quadrupole moment tensor in the form
\bea
{\cal D}^{\alpha\beta} = M_\odot R^2_\odot\sum_{n=-\infty}^\infty\,\sum_{m=-2}^2 C_m\,J_{nm}{\cal I}_{m}^{\alpha\beta}.
\nonumber
%\label{A11}
\ena
Inserting the above expansion into the right side of (\ref{A1}), taking into account (\ref{A6}), , we arrive at the following expression for the coefficients $C_m$ 
\bea
C_{0} = -\sqrt{5\over 4\pi},~~C_{m} = \sqrt{15\over 4\pi}
~~m=\pm1,\pm2.
\label{A12}
\ena

%%%%%%%%%%%%%%%%%%%%%%%%%%%%%%%%%%%%%%%%%%%%%%%%%%%%%%%%%%%%%%%%%%%%%%%%%%%%%%%%%%%%%%%%
%%%%%%%%%%%%%%%%%%%%%%%%%%%%%%%%%%%%%%%%%%%%%%%%%%%%%%%%%%%%%%%%%%%%%%%%%%%%%%%%%%%%%%%%

\section{ Derivation of the detector tensor $T_{\epsilon\gamma}$ \label{AppendixII}}

We use a cartesian coordinate system $x^\alpha$ with $x^\alpha x_\alpha = r^2$.
Then
\bea
\begin{array}{l}
\nabla_{\epsilon} \left(1\over r\right) = -{x_\epsilon\over r^3},
\\ \\
\nabla_{\gamma\epsilon} \left(1\over r\right)  =  -{\delta_{\epsilon\gamma}\over r^3}
+ 3{ x_\epsilon x_\gamma\over r^5},
\\ \\
\nabla_{\beta\gamma\epsilon} \left(1\over r\right)   = 
 +3{\delta_{\epsilon\gamma} x_\beta\over r^5}
+ 3{\delta_{\epsilon\beta} x_\gamma\over r^5}
+ 3{x_\epsilon \delta_{\gamma\beta}\over r^5}
-15 {x_\epsilon x_\gamma x_\beta\over r^7},
\\ \\
\nabla^4_{\alpha\beta\gamma\epsilon} \left(1\over r\right)  = 
 {3\over r^5}(\delta_{\epsilon\gamma} \delta_{\beta\alpha}
+ \delta_{\epsilon\beta} \delta_{\gamma\alpha}
+ \delta_{\epsilon\alpha} \delta_{\gamma\beta})\\ \\ 
~~~~~~~~~~\qquad - {15\over r^7}(\delta_{\epsilon\gamma} x_\beta x_\alpha
+ \delta_{\epsilon\beta} x_\gamma x_\alpha
+ \delta_{\epsilon\alpha} x_\gamma x_\beta
+ \delta_{\gamma\beta} x_\epsilon x_\alpha
+ \delta_{\gamma\alpha} x_\epsilon x_\beta
+ \delta_{\beta\alpha} x_\epsilon x_\gamma)
+ {105\over r^9} x_\epsilon x_\gamma x_\beta x_\alpha .
\end{array}
\label{B1}
\ena
Defining the unit vector in the radial direction
\bea
\bar{n}_\epsilon = \left({x_1\over r},\,{x_2\over r},\,{x_3\over r}\right),
\nonumber
\ena
we get
\bea
\nabla^4_{\alpha\beta\gamma\epsilon} \left(1\over r\right) =
&& {1\over r^5}{\bigg(}\,3\,(\delta_{\epsilon\gamma} \delta_{\beta\alpha}
+ \delta_{\epsilon\beta} \delta_{\gamma\alpha}
+ \delta_{\epsilon\alpha} \delta_{\gamma\beta})
\nonumber \\ && 
- 15\,(\delta_{\epsilon\gamma}\bar{n}_\beta \bar{n}_\alpha
+ \delta_{\epsilon\beta}\bar{n}_\gamma \bar{n}_\alpha
+ \delta_{\epsilon\alpha}\bar{n}_\gamma \bar{n}_\beta
+ \delta_{\gamma\beta}\bar{n}_\epsilon \bar{n}_\alpha
+ \delta_{\gamma\alpha}\bar{n}_\epsilon \bar{n}_\beta
+ \delta_{\beta\alpha}\bar{n}_\epsilon \bar{n}_\gamma)
 + 105\,\bar{n}_\epsilon \bar{n}_\gamma \bar{n}_\beta \bar{n}_\alpha {\bigg)}.
\nonumber 
%\label{B2}
\ena
Defining $\mu = n^\epsilon \bar{n}_\epsilon$ and contracting with $n^\epsilon n^\gamma$ gives
\bea
n^\epsilon n^\gamma \nabla^4_{\alpha\beta\gamma\epsilon} \left(1\over r\right) =
&& {3\over r^5}{\bigg(}\,\delta_{\alpha\beta}
+ n_{\beta} n_{\alpha}
+ n_{\alpha} n_{\beta} \nonumber \\ && - 5\,(\bar{n}_\beta \bar{n}_\alpha
+ \mu n_\beta \bar{n}_\alpha
+ \mu n_\alpha \bar{n}_\beta
+ \mu n_\beta \bar{n}_\alpha
+ \mu n_\alpha \bar{n}_\beta
+ \mu^2 \delta_{\alpha\beta})
+ 35\,\mu^2 \bar{n}_\beta \bar{n}_\alpha {\bigg)}.
\label{B3}
\ena
Since this tensor is to be contracted with the symmetric trace free tensor ${\cal D}^{\beta\alpha} = {\cal D}^{\alpha\beta}$  and $\delta_{\beta\alpha} {\cal D}^{\beta\alpha} = 0$ we only require the trace free components of this tensor. For this reason the terms with $\delta_{\alpha \beta}$ can be removed in (\ref{B3}) leaving
\bea
T_{\alpha\beta} = {r^5\over 3}\left[n^\epsilon n^\gamma  \nabla^4_{\alpha\beta\gamma\delta} \left(1\over r\right)\right]_{trace-free} =
2 n_{\beta} n_{\alpha} - 10\mu n_\beta \bar{n}_\alpha
- 10\mu n_\alpha \bar{n}_\beta + 5 (7 \mu^2 - 1) \bar{n}_\beta \bar{n}_\alpha.
\label{B4}
\ena

%%%%%%%%%%%%%%%%%%%%%%%%%%%%%%%%%%%%%%%%%%%%%%%%%%%%%%%%%%%%%%%%%%%%%%%%%%%%%%%%%%%%%%%%
%%%%%%%%%%%%%%%%%%%%%%%%%%%%%%%%%%%%%%%%%%%%%%%%%%%%%%%%%%%%%%%%%%%%%%%%%%%%%%%%%%%%%%%%

\section{The functions $f_j^{N}(\phi)$, $f_j^{GW}(\phi)$ \label{AppendixIII}}

Here we compute the source tensor in local transverse Cartesian coordinates $\xi^\alpha$ at a point $P$ on the orbit of the detector where $\xi^1,\xi^2$ are in the orbit plane, $\xi^1$ in the the outward radial direction, $\xi^2$ is in the direction of motion and $\xi^3$ perpendicular to the orbit plane. $\chi$ is the angle between $\xi^1$ and the $x^1$ direction of the coordinate system $x^\alpha$ in which the multipole moments are determined. The unit vectors along the $\xi^\alpha$ axes in the $x^\alpha$ coordinate system are
\bea
e_\xi^1 = (\cos\chi, \sin\chi, 0),~~
e_\xi^2 = (-\sin\chi, \cos\chi, 0),~~
e_\xi^3 = (0, 0, 1).
\nonumber
\ena
Hence the basis quadrupole tensors in the $\xi$ coordinate system are given by ${\sl I}_m^{\gamma\epsilon} = e^\gamma_\alpha e^\epsilon_\beta {\cal I}^{\alpha\beta}$ and are explicitly given by
\bea
\begin{array}{c}
{\sl I}_0=\left(\begin{array}{ccc} 1&0&0\\ 0&1&0\\ 0&0&-2\end{array}\right),~
{\sl I}_1=\left(\begin{array}{ccc} 0&0&\cos\chi\\ 0&0&-\sin\chi\\ \cos\chi&-\sin\chi&0\end{array}\right),~
{\sl I}_{-1}=\left(\begin{array}{ccc} 0&0&\sin\chi\\ 0&0&\cos\chi\\ \sin\chi&\cos\chi&0\end{array}\right),
\\ \\
{\sl I}_2=\left(\begin{array}{ccc} \cos2\chi&-\sin2\chi&0\\ -\sin2\chi&-\cos2\chi&0\\ 0&0&0\end{array}\right),~
{\sl I}_{-2}=\left(\begin{array}{ccc} \sin2\chi&\cos2\chi&0\\ \cos2\chi&-\sin2\chi&0\\ 0&0&0\end{array}\right).
\end{array}
\nonumber
\ena
The transverse-trace-free radiation tensors in the $\xi$ coordinates are then
\bea
\begin{array}{c}
\tilde{\sl I}_0={3\over 2} \left(\begin{array}{ccc}  0&0&0\\ 0&1&0\\ 0&0&-1 \end{array}\right),~
\tilde{\sl I}_1=-\sin\chi\left(\begin{array}{ccc} 0&0&0\\  0&0&1\\ 0&1&0\end{array}\right),~
\tilde{\sl I}_{-1}=\cos\chi \left(\begin{array}{ccc} 0&0&0\\ 0&0&1\\ 0&1&0 \end{array}\right), \nonumber \\
\\ \\
\tilde{\sl I}_2=-{1\over 2}\cos2\chi \left(\begin{array}{ccc}  0&0&0\\ 0&1&0\\ 0&0&-1\end{array}\right),~
\tilde{\sl I}_{-2}=-{1\over 2}\sin2\chi \left(\begin{array}{ccc}  0&0&0\\ 0&1&0\\ 0&0&-1\end{array}\right).
\end{array}
\nonumber
\ena
Now let the plane of the detector be at an inclination angle $\iota$ to the plane of the orbit ($\iota = \pi/3$ in the LISA experiment), and let the detector arm AB be at an angle $\psi$ to the direction of the orbit ($\xi^2$). In the the $\xi$ coordinate system the the unit vector along the arm of the arm AB of the detector is
\bea
n^\alpha = (-\sin\psi\cos\iota, \cos\psi, \sin\psi\sin\iota),
\nonumber 
\ena
so the projection tensor $N_{AB}^{\alpha\beta} = n^\alpha n^\beta$ is
\bea
N_{AB}=\left(\begin{array}{ccc} \sin^2\psi\cos^2\iota&-\sin\psi\cos\psi\cos\iota&
-\sin^2\psi\sin\iota\cos\iota\\ -\sin\psi\cos\psi\cos\iota&\cos^2\psi&\cos\psi\sin\psi\sin\iota\\ -\sin^2\psi\sin\iota\cos\iota&\cos\psi\sin\psi\sin\iota&\sin^2\psi\sin^2\iota
\end{array}\right).
\nonumber
\ena
Let the angle between the arms AC and AB of the detector be $\upsilon$ so that AC makes an angle $\psi + \upsilon$ with the $\xi^2$ direction, the projection tensor $N_{AC}^{\alpha\beta}$ is given by the above result with $\psi$ replaced by $\psi + \upsilon$ so the difference
\bea
\Delta N = N_{AB} - N_{AC} =
\sin\upsilon\left(
\begin{array}{ccc}
-\sin(2\psi+\upsilon)\cos^2\iota&\cos(2\psi+\upsilon)\cos\iota&
\sin(2\psi+\upsilon)\sin\iota\cos\iota\\ \cos(2\psi+\upsilon)\cos\iota&
\sin(2\psi+\upsilon)&-\cos(2\psi+\upsilon)\sin\iota\\ \sin(2\psi+\upsilon)\sin\iota\cos\iota&
-\cos(2\psi+\upsilon)\sin\iota&-\sin(2\psi+\upsilon)\sin^2\iota
\end{array}\right).
\nonumber
\ena
The signal functions for gravitational waves $f_j^{GW}$ are given by
\bea
f^{GW}_m = \Delta N_{\alpha\beta} \tilde{\sl I}_m^{\alpha\beta} =
\sin\upsilon\left(\sin(2\psi+\upsilon)\,\tilde{\sl I}_m^{22} -2\cos(2\psi+\upsilon)\sin\iota\,\tilde{\sl I}_m^{23} -
\sin(2\psi+\upsilon)\sin^2\iota\,\tilde{\sl I}_m^{33}\right),
\nonumber
\ena
which gives
\bea
\begin{array}{l}
f^{GW}_0 = ~~1.5\,\sin\upsilon\,(1+\sin^2\iota)\,\sin(2\psi+\upsilon),
\\ \\
f^{GW}_1 = ~~2\,\sin\upsilon\,\sin\iota\,\sin\chi\,\cos(2\psi+\upsilon),
\\ \\
f^{GW}_{-1} = -2\,\sin\upsilon\,\sin\iota\,\cos\chi\,\cos(2\psi+\upsilon),
\\ \\
f^{GW}_{2} = -0.5\,\sin\upsilon\,(1+\sin^2\iota)\,\cos{2\chi}\,\sin(2\psi+\upsilon),
\\ \\
f^{GW}_{-2} = -0.5\,\sin\upsilon\,(1+\sin^2\iota)\,\sin{2\chi}\,\sin(2\psi+\upsilon)).
\end{array}
\label{C1}
\ena
The Newtonian signal functions $f^{N}_m=\Delta T_{\alpha\beta} {\sl I}_m^{\alpha\beta}$ where
\bea 
\Delta T_{\alpha\beta} = 2 \Delta N_{\alpha\beta} + 10 \Delta M_{\alpha\beta}
+ 5 \Delta R_{\alpha\beta},
\nonumber
\ena
with
\bea
M_{\alpha\beta} = \mu(\bar{n}^\alpha n^\beta + \bar{n}^\beta n^\alpha)
,~~~R_{\alpha\beta} = (7\mu^2  - 1) \bar{n}^\alpha\bar{n}^\beta,~~\mu = \bar{n}^\gamma n_\gamma.
\nonumber
\ena
In the $\xi$ coordinates $\bar{n}^\alpha = (1,0,0)$, and for the detector arm AB $\mu = n^\alpha \bar{n}_\alpha = -\sin\psi \cos\iota$ so
\bea
M_{AB}=
\left(\begin{array}{ccc}
2\sin^2\psi\cos^2\iota&-\sin\psi\cos\psi\cos\iota&
-\sin^2\psi\sin\iota\cos\iota\\ -\sin\psi\cos\psi\cos\iota&0&0\\ -\sin^2\psi\sin\iota\cos\iota&0&0
\end{array}\right),
\nonumber
\ena
and hence
\bea
\Delta M = M_{AB} - M_{AC} =
\sin\upsilon
\left(\begin{array}{ccc}
-2\sin(2\psi+\upsilon)\cos^2\iota&\cos(2\psi+\upsilon)\cos\iota&
\sin(2\psi+\upsilon)\sin\iota\cos\iota\\ \cos(2\psi+\upsilon)\cos\iota&0&0\\ \sin(2\psi+\upsilon)\sin\iota\cos\iota&0&0
\end{array}\right).
\nonumber
\ena
The tensor $R_{AB} = (7\mu^2-1)\bar{n}^\alpha \bar{n}^\beta$ gives
\bea
\Delta R = -7 \sin\upsilon \sin(2\psi+\upsilon) \cos^2\iota
\left(\begin{array}{ccc}
1&0&0\cr 0&0&0\cr 0&0&0
\end{array}\right).
\nonumber
\ena
So finally the tensor $\Delta T_{\alpha\beta}$ is
\bea
\Delta T =
\sin\upsilon
\left(\begin{array}{ccc}
-17\sin(2\psi+\upsilon)\cos^2\iota&-8 \cos(2\psi+\upsilon)\cos\iota&
-8\sin(2\psi+\upsilon)\sin\iota\cos\iota\cr -8\cos(2\psi+\upsilon)\cos\iota& 2 sin(2\psi+\upsilon)&-2 \cos(2\psi+\upsilon)\sin\iota\cr -8 \sin(2\psi+\upsilon)\sin\iota\cos\iota&-2\cos(2\psi+\upsilon)\sin\iota&-2\sin(2\psi+\upsilon)\sin^2\iota
\end{array}\right).
\nonumber
\ena
The signal functions $f^N_m= {\sl I}_m^{\alpha\beta} \Delta T_{\alpha\beta}$ are then
\bea
\begin{array}{l}
f^{N}_0 =\sin\upsilon\,(6-21\cos^2\iota)\,\sin(2\psi+\upsilon),
\\ \\
f^N_1 = -4\,\sin\upsilon\,\sin\iota\,\left[4\,\cos\iota\,\cos\chi\,\sin(2\psi+\upsilon)
-\sin\chi\,\cos(2\psi+\upsilon)\right],
\\ \\
f^N_{-1} = -4\,\sin\upsilon\,\sin\iota\,\left[4\,\cos\iota\,\sin\chi\,
\sin(2\psi+\upsilon) + \cos\chi\,\cos(2\psi+\upsilon)\right],
\\ \\
f^N_2 = -\sin\upsilon\,\left[(17\cos^2\iota +2)\,\cos{2\chi}\,\sin(2\psi+\upsilon) -
16\,\cos\iota\,\sin{2\chi}\,\cos(2\psi+\upsilon)\right],
\\ \\
f^N_{-2} = -\sin\upsilon\,\left[(17\cos^2\iota +2)\,\sin{2\chi}\,\sin(2\psi+\upsilon) +
16\,\cos\iota\,\cos{2\chi}\,\cos(2\psi+\upsilon)\right].
\end{array}
\label{C2}
\ena
Setting $\iota=\upsilon=\pi/3$ in (\ref{C1}) and (\ref{C2}) brings us to expression (\ref{30}) for the specific case of LISA.

%%%%%%%%%%%%%%%%%%%%%%%%%%%%%%%%%%%%%%%%%%%%%%%%%%%%%%%%%%%%%%%%%%%%%%%%%%%%%%%%%%%%%%%%
%%%%%%%%%%%%%%%%%%%%%%%%%%%%%%%%%%%%%%%%%%%%%%%%%%%%%%%%%%%%%%%%%%%%%%%%%%%%%%%%%%%%%%%%

%%%%%%%%%%%%%%%%%%%%%%%%%%%%%%%%%%%%%%%%%%%%%%%%%%%%%%%%%%%%%%%%%%%%%%%%%%%%%%%%%%%%%%%%
%%%%%%%%%%%%%%%%%%%%%%%%%%%%%%%%%%%%%%%%%%%%%%%%%%%%%%%%%%%%%%%%%%%%%%%%%%%%%%%%%%%%%%%%

\end{document}